\documentclass[a4paper,11pt]{article}
\pdfoutput=1 % if your are submitting a pdflatex (i.e. if you have
\usepackage{jheppub} % for details on the use of the package, please
\usepackage[T1]{fontenc} % if needed

\title{Bottom-charmed meson states in inverse problem of QCD}

\author[a]{Halil Mutuk}
\author[b]{Duygu Yıldırım}

\affiliation[a]{Department of Physics, Faculty of Sciences, Ondokuz Mayis University, 55139 Samsun, Türkiye}
\affiliation[b]{Department of Physics, Faculty of Arts and Sciences, Amasya University,  05200 Amasya, Türkiye}

\emailAdd{hmutuk@omu.edu.tr}
\emailAdd{yildirimyilmaz@amasya.edu.tr}

\abstract{We present a comprehensive analysis of the bottom–charmed ($B_c$) meson spectrum within the inverse matrix QCD sum rules formalism. In this framework, conventional QCD sum rules are recast as an inverse problem, allowing for the direct reconstruction of hadronic spectral densities from first principles without invoking phenomenological continuum parametrizations or quark–hadron duality assumptions. We compute the masses and decay constants of conventional $B_c$ mesons with quantum numbers $J^P = 0^-$, $1^-$, $0^+$, and $1^+$. The obtained results are in close agreement with available experimental measurements and are consistent with predictions from various theoretical and phenomenological approaches. The inverse matrix formulation exhibits improved numerical stability and reduced systematic uncertainties relative to standard implementations, highlighting its suitability for precision spectroscopy of heavy quarkonium systems.}

\begin{document} 
\maketitle
\flushbottom

%%%%%%%%%%%%%%%%%%%%%%%%%%%%
\section{Motivation}
\label{sec:intro}
%%%%%%%%%%%%%%%%%%%%%%%%%%%%

The quest to understand the strong interaction, governed by Quantum Chromodynamics (QCD), represents a central frontier in theoretical particle physics. While the Lagrangian of QCD is elegantly simple, its nonperturbative nature at low energies gives rise to the profound complexity of the hadronic spectrum. Heavy quarkonium systems—bound states of a heavy quark and its antiquark—serve as a privileged laboratory in this endeavor. Their relative simplicity, owing to the large quark masses which justify a nonrelativistic treatment, allows them to act as a ``hydrogen atom'' for QCD, providing a testing ground for our computational techniques. The spectroscopy and decay properties of heavy quarkonium systems, such as charmonium $(c \bar{c})$ and bottomonium $(b \bar{b})$, have served as fundamental testing grounds for QCD. These systems provide critical insights into the interplay between perturbative and nonperturbative phenomena in the strong interaction. 

The study of heavy mesons has long provided an essential testing ground for our understanding of QCD. Among these systems, the $B_c$ meson is particularly distinctive. As a bound state of a bottom quark and a charm antiquark $(b \bar{c})$ (or vice versa) it represents the only known meson composed of two different heavy flavors. This unique structure not only differentiates it from the better-studied charmonium $(c \bar{c})$ and bottomonium $(b \bar{b})$ systems but also grants it several special properties. The quarkonium family is traditionally divided into $(c \bar{c})$ and $(b \bar{b})$. These states, being composed of a quark and its own antiquark, are eigenstates of the charge conjugation operation. A crucial feature of their phenomenology is the presence of annihilation decays into gluons or photons. However, the $B_c$ meson is completely different. As a flavored, non-self-conjugate meson, it is the lowest-lying bound state with two different heavy flavors, ``open'' in both flavor sectors. Notwithstanding, the $B_c$ meson cannot annihilate into gluons or photons, and its ground state decays exclusively via the weak interaction. It is the only experimentally observed doubly heavy meson that is stable under the strong interaction. These characteristics make $B_c$ mesons a privileged laboratory for investigating the interplay of strong, weak, and even electromagnetic forces within a single hadronic system.

The first experimental evidence for the pseudoscalar $B_c$ meson was obtained in 1998 by the CDF collaboration at Fermilab, through the observation of semileptonic decays in $p \bar{p}$ collisions at $\sqrt{s}=1.8 \ \text{TeV}$ with a measured mass of $M_{B_c}=6.40 \pm 0.39 \pm 0.13 \ \text{GeV}$ \cite{CDF:1998axz,CDF:1998ihx}. This marked the first confirmed identification of a bottom–charm bound state. Later measurements consolidated these findings: in 2007 the CDF collaboration reconstructed non-leptonic decays $B_c^{\pm} \to J/\psi \pi^{\pm}$  with the measured mass $M_{B_c}=6275.6 \pm 2.9\pm 2.5 \ \text{MeV}$, while in 2008 the D0 collaboration reported further confirmation of its mass as $M_{B_c}=6300 \pm 14 \pm 5\ \text{ MeV}$. This state is now well established in the Particle Data Group (PDG) \cite{ParticleDataGroup:2024cfk} with a mass of $M_{B_c}=6274.47\pm 0.27\pm 0.17\,\rm{MeV}$. 

The ongoing experiments have revealed excited $B_c$ states. The invariant mass spectrum of $B_c^\pm \pi^+\pi^-$ combinations was analyzed by the ATLAS collaboration in 2014, leading to the discovery of a new structure. With an observed mass of $M=6842\pm4\pm 5 \text{MeV}$ and a statistical significance exceeding five standard deviations, the observed mass and decay characteristics of this particular state align well with the theoretical predictions for the $B_c^{\pm}(2S)$ meson, which is the second S-wave excitation of the $B_c^{\pm}$ meson \cite{ATLAS:2014lga}. The CMS collaboration reported in 2019 the observation of two distinct excited $(\bar{b}c)$ states \cite{CMS:2019uhm}. These states were detected as structures in the $B^+_c\pi^+\pi^-$  invariant mass spectrum, with a statistical significance greater than five standard deviations $(5\sigma)$. These two signals are consistent with the $B_c^+(2S)$ and $B_c^{*+}(2S)$ states, respectively. The two peaks are well resolved, with a measured mass difference of $\Delta M = 29.1 \pm 1.5 \, (\text{stat}) \pm 0.7 \, (\text{syst}) \, \text{MeV}$. The $B_c^+ (2S)$ mass is measured to be $M_{B_c^+ (2S)}=6871.0 \pm 1.2 \, (\text{stat}) \pm 0.8 \, (\text{syst}) \pm 0.8 \, (B_c^+) \, \text{MeV}$, where the last term is the uncertainty in the world-average $B_c^+$ mass. The observed peak mass for the $B_c^{*+}(2S)$ is a lower limit, as the true value is obscured due to the experimental inability to reconstruct the low-energy photon emitted during the $B_c^{*+} \to B_c^+ \gamma$ radiative decay. In the same year, LHCb collaboration observed a peaking structure in the $B_c^+ \pi^+ \pi^-$ mass spectrum (with a global (local) statistical significance of $6.3\sigma$ ($6.8\sigma$)) which is consistent with the $B_c^{*+}(2S)$ state with a mass of $M_{B_c^{*+}(2S)}=6841.2 \pm 0.6  \pm 0.1  \pm 0.8 \,\text{MeV}$. The data also contain a marginal indication of a second structure, which aligns with the expected $B_c^{+}(2S)$ state. This tentative structure has a global (local) statistical significance of $2.2\sigma$ ($3.2\sigma$). If this signal is attributed to the $B_c^+(2S)$ meson, its mass is determined to be $M_{B_c^+(2S)}=  6872.1 \pm 1.3  \pm 0.1  \pm 0.8 \,\text{MeV}$ \cite{LHCb:2019bem}. 

The first observation of the orbitally excited $B_c(1P)^+$ states by the LHCb collaboration in 2025 represented a major advancement in beauty-charm spectroscopy~\cite{LHCb:2025uce}. Analyzing $9~\text{fb}^{-1}$ of $pp$ collision data, a significant structure ($>7\sigma$) was observed in the $B_c^+\gamma$ mass spectrum. The data are best described by an effective model of two unresolved peaks at
$M_1 = 6704.8 \pm 5.5 \pm 2.8 \pm 0.3~\text{MeV} $ and $M_2 = 6752.4 \pm 9.5 \pm 3.1 \pm 0.3~\text{MeV}$, interpreted as the overlapping signals of the four predicted $B_c(1P)^+$ states. This analysis reports a structure in the $B_c^+\gamma$ mass spectrum whose features are compatible with the predicted $B_c(1P)^+$ states, however since the resolution is insufficient to resolve the fine structure of the $B_c(1P)^+$ multiplet, it prevents the clear identification of the individual $1^3P_0$, $1P_1$, $1P_1'$, and $1^3P_2$ states. Furthermore, the measured peak locations corroborate theoretical models, signifying the first observation of orbitally excited $B_c$ mesons. A companion analysis measured the relative production rate $R = \sigma(B_c(1P)^+)/\sigma(B_c^+) = 0.20 \pm 0.03 \pm 0.02 \pm 0.03$, consistent with nonrelativistic QCD (NRQCD) predictions \cite{LHCb:2025ubr}. While this marks a discovery, the individual $1P$ states remain unresolved, and precise determinations of their masses, spin-parity, and the $B_c^{*+}$-$B_c^+$ mass splitting are still lacking.

Spectroscopic parameters (such as mass and decay constant) of $B_c$ mesons are studied with using various models such as relativistic quark potential model \cite{Godfrey:1985xj,Zeng:1994vj,Gupta:1995ps,Ebert:2002pp,Godfrey:2004ya}, nonrelativistic quark model \cite{Gershtein:1994dxw,Eichten:1994gt,Fulcher:1998ka,Ebert:2011jc,Monteiro:2016ijw,Soni:2017wvy,Eichten:2019gig,Li:2019tbn,Ortega:2020uvc}, covariant light-front quark model \cite{Verma:2011yw,Tang:2018myz}, shifted large-N expansion \cite{Ikhdair:2003ry,Ikhdair:2006nx}, perturbative QCD \cite{Brambilla:2000db}, nonrelativistic renormalization group \cite{Penin:2004xi}, lattice QCD \cite{Davies:1996gi,Jones:1998ub,Gregory:2009hq,Mathur:2018epb}, QCD sum rules (QCDSR) \cite{Colangelo:1992cx,Chabab:1993nz,Kiselev:1993ea,Kiselev:1995bv,Bagan:1994dy,Kiselev:2003uk,Wang:2012kw,Wang:2013cdy,Baker:2013mwa,Aliev:2019wcm,Narison:2019tym,Wang:2024fwc}, light-cone QCDSR \cite{Ozdem:2024qaa}, heavy quark effective theory \cite{Onishchenko:2003ui,Lee:2010ts,Chen:2015csa,Tao:2022qxa,Tao:2022hos,Sang:2022tnh,Feng:2022ruy,Tao:2023pzv}, Bethe-Salpeter equation \cite{AbdEl-Hady:1998uiq,Wang:2007av,Wang:2022cxy} and field correlator method \cite{Badalian:2007km}. However, the understanding of the $B_c$ meson is surprisingly limited. There are just two states of bottom-charmed meson listed in the PDG \cite{ParticleDataGroup:2024cfk}. This lack of information stands in sharp contrast to the well-developed and detailed spectra of similar particles like charmonium and bottomonium, whose many excited states are well-established. Despite recent progress in heavy quark physics, the full family of $B_c$ states remains poorly mapped. Therefore, extensive further investigation is essential to complete our knowledge of this system.

The exploration of the mass spectrum and decay constants of mesons is fundamental for advancing our understanding of both strong and weak interactions. The mass spectrum of mesons reflects the dynamics of quark–antiquark binding under the strong force and provides a direct window into the nonperturbative regime of QCD. For any meson, the spectrum of states provides a direct map of the confining QCD potential, with different energy levels probing its short-distance Coulombic and long-range linear regimes. It serves as the critical benchmark for ab-initio calculations like lattice QCD and for effective theories such as NRQCD, testing their ability to describe the interplay of two different mass scales within a single bound state. Precise measurements and theoretical predictions of meson masses allow us to rigorously test different frameworks, such as potential models, lattice QCD, Bethe–Salpeter equations, and QCDSR formalism. Any discrepancy between experiment and theory may signal the need for improved treatments of heavy-quark dynamics or, potentially, new physics beyond the Standard Model. In the case of heavy–heavy systems like the $B_c$ meson, the spectrum is especially informative because it tests aspects of heavy-quark effective theory and provides unique insights not accessible in charmonium or bottomonium. Equally important are decay constants, which quantify the overlap of the meson internal wave function and parameterize the coupling of the meson to weak currents. They enter directly into the calculation of purely leptonic decays, where the decay rate is proportional to the square of the decay constant. Beyond leptonic processes, decay constants also serve as essential inputs for determining form factors, partial decay widths, and branching fractions in semileptonic and hadronic decays.  In particular, decay constants enter the theoretical description of distribution amplitudes, form factors, partial decay widths, and branching fractions. Most importantly, decay constants are indispensable for extracting Cabibbo–Kobayashi–Maskawa (CKM) matrix elements from experimental data. Precise determinations of CKM elements, such as $V_{cb}$ are crucial for testing the unitarity of the CKM matrix and for probing possible contributions from new physics. 

Despite substantial progress, results for the mass spectrum and decay constants obtained from different theoretical frameworks often show significant variation and, in some cases, mutual inconsistency. Predictions from quark models, lattice QCD, Bethe–Salpeter approaches, and QCDSR can differ by several hundred MeV for the mass spectrum and by large margins for decay constants. These discrepancies highlight the limitations of current methods and the sensitivity of predictions to approximations such as relativistic corrections, choice of potential, or truncations in operator expansions. Consequently, there is a  need to either develop new methods or refine existing approaches by incorporating higher-order corrections, improved parameter inputs, and better treatments of nonperturbative dynamics. Such efforts are essential to achieve reliable and precise predictions that can be meaningfully compared with the increasingly accurate data from ongoing and future collider experiments.

In this context, one promising refinement is the inverse matrix QCDSR formalism, which improves upon traditional QCDSR by stabilizing the extraction of hadronic parameters and reducing sensitivity to uncertainties in the choice of Borel windows or continuum thresholds. By reorganizing the sum rule equations into a matrix structure and solving them inversely, this method provides more reliable determinations of meson masses and decay constants. At the same time, the validity of these approaches relies on the principle of quark–hadron duality, which bridges the quark–gluon description of correlation functions at short distances with hadronic observables at long distances. Improving how duality violations are treated, or reducing their impact through refined sum rule techniques, is critical for achieving consistency among theoretical models and with experimental data.

This paper is organized as follows. In Section~\ref{sec:qcdform}, we provide a brief overview of the conventional QCDSR formalism. Section~\ref{sec:qcdinverse} is dedicated to the theoretical foundation of the inverse matrix method, detailing its formulation as a solution to an inverse problem. Our numerical analysis for the masses and decay constants of the scalar, pseudoscalar, vector, and axialvector $B_c$ mesons is presented and discussed in Section~\ref{sec:numres}. Finally, we summarize our findings and offer concluding remarks in Section~\ref{conclusion}. To the best of our knowledge, this work represents the first comprehensive application of the inverse matrix QCDSR formalism to the complete spectrum of $B_c$ mesons.

%%%%%%%%%%%%%%%%%%%%%%%%%%%%
\section{QCDSR formalism}
\label{sec:qcdform}
%%%%%%%%%%%%%%%%%%%%%%%%%%%%

The QCDSR approach, first developed by Shifman, Vainshtein, and Zakharov  \cite{Shifman:1978bx,Shifman:1978by}, is a powerful nonperturbative tool for studying hadron properties by connecting QCD at the quark–gluon level to measurable hadronic observables. Its core philosophy is to relate the properties of a specific hadron to the vacuum structure of QCD through the use of correlation functions and the Operator Product Expansion (OPE). The starting point is the computation of a two-point correlation function in the deep Euclidean momentum region $(Q^2=-q^2 \gg 0)$ where perturbative QCD is applicable. For a generic interpolating current $J(x)$ with the quantum numbers of the hadron of interest, the correlation function is defined as
\begin{equation}
\Pi(q^2) = i \int d^4x \, e^{iq\cdot x} \langle 0|T\{J(x)J^\dagger(0)\}|0\rangle, \label{corfunc}
\end{equation}
where $T$ denotes the time-ordered product. At large spacelike momentum transfer, this correlation function can be expanded using the OPE, which systematically separates short-distance perturbative effects from long-distance nonperturbative contributions encoded in vacuum condensates. On the other hand, the same correlation function can be expressed at the hadronic level through a dispersion relation involving physical spectral densities.

On the theoretical side, this correlator is evaluated using the OPE, which factorizes the short- and long-distance dynamics
\begin{equation}
\Pi^{\rm OPE} (q^2)=\sum_d C_d (q^2) \langle \hat{O}_d \rangle \label{corqcd}.
\end{equation}
Here, the coefficient functions $C_d (q^2)$ —the Wilson coefficients—are process-dependent and calculable in perturbative QCD. The local gauge-invariant operators $\hat{O}_d $ represent the nonperturbative QCD vacuum condensates. Simultaneously, the correlation function is expressed through a dispersion relation in its phenomenological representation, which is saturated by a spectrum of physical states
\begin{equation}
\Pi^{\text{(Phen)}}(q^2) = \frac{\langle 0|J|H\rangle \langle H|J^\dagger|0\rangle}{m_H^2 - q^2} + \cdots, \label{corphen}
\end{equation}
where the first term represents the ground state hadron $H$ of mass $m_H$, and $\cdots$ denote the contribution of higher resonances and continuum states. The coupling of the interpolating current to the physical hadron is quantified by the decay constant, which parameterizes the corresponding hadronic matrix element
\begin{equation}
\langle 0|J|H\rangle= \lambda.
\end{equation}
With this definition Eq. (\ref{corphen}) can be written as
\begin{equation}
\Pi^{\text{(Phen)}}(q^2)=\frac{\lambda^2}{m_H^2 - q^2} + \cdots . \label{corphen1}
\end{equation}
An alternative way to write $\Pi^{\text{(Phen)}}(q^2)$ is to use dispersion integral in terms of the spectral density $\rho^{(\rm Phen)}(s)$ as
\begin{equation}
\rho^{(\rm Phen)}(s)=\lambda^2 \delta (s-m_H^2) + \rho^h(s) \theta (s-s_0),
\end{equation}
where $\lambda^2 \delta (s-m_H^2)$ represents the pole term, $s_0$ is the continuum threshold parameter, $ \theta (s-s_0)$ is the step function, and $ \rho^h(s)$ is the unknown hadronic spectral density. After this consideration, phenomenological representation of the correlation function can be written as
\begin{equation}
\Pi^{\text{(Phen)}}(q^2) = \frac{\lambda^2}{m_H^2 - q^2} + \frac{1}{\pi} \int_{s_0}^\infty ds \frac{\rho^h(s)}{s - q^2}. \label{corphen2}
\end{equation}

In terms of dispersion relation, it is possible to write Eq. (\ref{corqcd}) as
\begin{equation}
\Pi^{\rm OPE}(q^2)=\int_0^{s_0} \frac{\rho^{\rm OPE} (s)}{s-q^2} ds, \label{opespec}
\end{equation}
where the spectral density can be written by $\rho^{\rm OPE} (s)=\frac{1}{\pi} \rm{Im} \Pi (s)$. The spectral density $\rho (s)$ is often modeled using the quark-hadron duality assumption, which posits that above $s_0$ the spectral density coincides with the perturbative QCD prediction
\begin{equation}
\rho^h(s) \simeq \rho^{\rm OPE}(s), \label{qhdual}
\end{equation}
for $s > s_0$. 

The final step is to match the two representations, $\Pi^{\rm OPE}(q^2)$  and $\Pi^{\rm Phen}(q^2)$, i.e. equating Eq. (\ref{corqcd}) and Eq. (\ref{corphen}). To improve the convergence of the OPE and enhance the contribution of the ground state, a Borel transform $\hat{B}$ can be applied
\begin{equation}
\mathcal{B}[\Pi(q^2)] = \lim_{{(-q^2)^n \to \infty}} \frac{(-q^2)^n}{(n-1)!}\left(\frac{d}{dq^2}\right)^n \Pi(q^2). \label{borelt}
\end{equation}
This transformation suppresses the continuum and higher-dimensional condensate terms, leading to the master equation of the QCDSR. In the $q^2 \ll 0$ region, applying Borel transformation to phenomenological side of the correlation function yields
\begin{equation}
\mathcal{B}[\Pi^{\text{(Phen)}}(q^2)]= \lambda^2 e^{-m_H/M^2} + \int_{s_0}^\infty \rho^h(s) e^{-s/M^2}ds, \label{boreltphen}
\end{equation}
where $M^2$ is the Borel parameter and $s_0$ is the continuum threshold. It is also possible to apply Borel transformation to $\Pi^{\rm OPE}(q^2)$ as
\begin{equation}
\mathcal{B}[\Pi^{\text{(OPE)}}(q^2)]=\int_0^{s_0} \rho^{\rm OPE} (s) e^{-s/M^2} ds.
\end{equation}

The formalism proceeds by constructing a dispersion relation for the OPE amplitude $\Pi ^{\mathrm{OPE}}(q^{2})$ using the corresponding spectral density $\rho ^{\mathrm{OPE}}(s)$. The fundamental step of the sum rule derivation is to equate the Borel transforms of the physical representation, $\Pi ^{\mathrm{Phen}}(q^{2})$, and its OPE counterpart $\Pi^{\rm OPE}(q^2)$. At this stage, the principle of quark-hadron duality is applied, which approximates the physical spectral density $\rho ^{\mathrm{h}}(s)$ by $\rho ^{\mathrm{OPE}}(s)$ for energies exceeding the continuum threshold $s_0$. This assumption enables the removal of the continuum contribution, corresponding to the second term in Eq. (\ref{boreltphen}), from the final expression and gives 
\begin{equation}
\lambda^2 e^{-m_H^2/M^2}=\int_0^{s_0}  \rho^{\rm OPE} (s) e^{-s/M^2} ds
\end{equation}
the final sum rule for the physical parameters. Defining $\Pi(M^2,s_0) \equiv \int_0^{s_0}  \rho^{\rm OPE} (s) e^{-s/M^2} ds$ would simplify mathematical manipulations to  extract the mass and decay constant of the ground-state hadron. Taking derivatives with respect to $(-1/M^2)$, one can obtain the sum rules for mass and decay constant as follows
\begin{eqnarray}
m_H^2 &=& \frac{\Pi^\prime(M^2,s_0)}{\Pi(M^2,s_0)}, \\
\lambda^2&=&e^{m_H^2/M^2}\Pi(M^2,s_0), \label{sumrules}
\end{eqnarray}
respectively, where $\Pi^\prime(M^2,s_0)=\frac{d}{d(-1/M^2)}\Pi(M^2,s_0)$. Analyzing these sum rules will yield numerical values for mass and decay constant of the related hadron. The current phase of the analysis mandates the rigorous determination of the working windows for the two phenomenologically nonphysical auxiliary parameters in the QCDSR formalism: the Borel mass parameter $(M^2)$ and the continuum threshold $(s_0)$. Although these variables do not correspond to physical observables, their permissible ranges must strictly conform to the inherent methodological self-consistency constraints imposed by the sum rule approach; consequently, their selection cannot be arbitrary. This systematic dependence on the stability parameters constitutes one of the principal sources of theoretical uncertainty within the QCDSR formalism. The remaining major sources of nonparametric uncertainty are categorized as follows \cite{Leinweber:1995fn}:
\begin{itemize}
\item OPE Truncation Error: The necessity of truncating the OPE at a finite order introduces systematic error, which must be assessed by ensuring the rapid suppression and subdominance of the highest dimension operator term included. 
\item Uncertainty in QCD Vacuum Condensates: The final result is highly sensitive to the precise, nonperturbative values assigned to the universal QCD input parameters, such as quark condensate $\langle q \bar q \rangle$ and gluon condensate $\langle \alpha_s/\pi G^2 \rangle $ whose values are determined phenomenologically or via lattice QCD.
\item Quark-Hadron Duality Violation: The fundamental assumption that the calculated OPE (QCD side) accurately models the integrated physical spectral density (hadronic side) up to the threshold $(s_0)$ is an approximation. The inherent degree of duality violation introduces an irreducible systematic uncertainty.
\item Hadronic Spectral Function Modeling: In channels where the physical spectral density is not directly known from experiment, the reliance on resonance-plus-continuum models, particularly the parametrization of the continuum, contributes significant systematic uncertainty.
\item Parametric Dependence on Fundamental Constants: Uncertainties in input quantities like the running strong coupling constant $(\alpha_s)$ and the current quark masses propagate directly to the final extracted hadronic property.
\end{itemize}

Among these sources of uncertainty, quark-hadron duality is also important as well as Borel and continuum threshold parameters. Quark-hadron duality is a fundamental, nonproven assumption in QCD that provides a crucial bridge between theoretical calculations involving quarks and gluons and experimental measurements involving hadrons. In essence, quark-hadron duality states that, when averaged over a sufficiently large energy range, the physical observable described by the sum over hadronic resonances (hadrons) is approximately equivalent (dual) to the same quantity calculated using the quark and gluon degrees of freedom via perturbative QCD and the OPE \cite{Poggio:1975af}.

The need for quark-hadron duality arises from the two distinct regimes of QCD:
\begin{itemize}
\item High-Energy (Short Distance): Due to Asymptotic Freedom, the strong coupling $\alpha_s$ is small. Calculations are reliably performed using perturbation theory in terms of quarks and gluons.
\item Low-Energy (Long Distance): Due to Confinement, the strong coupling is large. Quarks and gluons bind into hadrons. Calculations are nonperturbative and highly complex.
\end{itemize}

Quark-hadron duality allows to use the simpler, calculable quark-gluon description in the short-distance (high-energy) regime to approximate the complex, summed-over-hadrons description in the long-distance (low-energy) regime, provided the calculation is sufficiently inclusive (averaged over a range of energy).  QCDSR formalism relies directly on quark-hadron duality through a dispersion relation to calculate hadronic properties:
\begin{equation}
\int_{\text{threshold}}^{s_0} ds \cdot \text{Im}[\Pi^{\text{Hadron}}(s)] = \int_{\text{threshold}}^{s_0} ds \cdot \text{Im}[\Pi^{\text{OPE}}(s)].
\end{equation}

In the left-hand side (hadronic side), the integral of the physical spectral density, $\text{Im}[\Pi^{\text{Hadron}}(s)]$, is modeled as a sum over the contributions of the lowest-lying resonance(s) and a continuum of higher-mass states, up to a theoretical threshold, $s_0$. The right-hand side (OPE side), the integral of the theoretical spectral density, $\text{Im}[\Pi^{\text{OPE}}(s)]$, is calculated using the OPE, which includes perturbative terms and nonperturbative vacuum condensates. The assumption of quark-hadron duality is that these two integrals are approximately equal, allowing the hadronic parameters (like mass or decay constant) to be extracted from the QCD calculation. 

Although quark-hadron duality is widely used in QCDSR formalism as well as in QCD motivated models, there are some indications that quark-hadron duality may be violated especially at low energies or over small energy ranges \cite{Gonzalez-Alonso:2010kpl,Boito:2017cnp,Pich:2022tca,Mannel:2024crj}. QCDSR is based on the QCD Lagrangian and an analytically solvable model. Therefore it is a frequently used framework for studying both conventional and exotic hadrons and the predictions agree well with the experimental and theoretical results in the literature. However,  accuracy of the formalism is constrained to an uncertainty of approximately  20\% due to the necessity of incorporating phenomenological input parameters, auxiliary parameters and other assumptions mentioned above. Therefore any attempts for improving the method will help to understand both perturbative and nonperturbative QCD.

%%%%%%%%%%%%%%%%%%%%%%%%%%%%
\section{Inverse matrix QCDSR approach}
\label{sec:qcdinverse}
%%%%%%%%%%%%%%%%%%%%%%%%%%%%

QCDSR formalism constitutes one of the most successful nonperturbative techniques for relating hadronic observables to the underlying quark and gluon dynamics.  By combining dispersion relations, the OPE, and quark--hadron duality, the method provides a quantitative bridge between QCD and the hadronic spectrum. The conventional formulation relies on modeling the spectral density as a sum of a pole representing the lowest-lying resonance and a perturbative continuum beginning at an effective threshold $s_0$.  The Borel transformation is then employed to suppress subtraction terms and higher-state contributions.  Although this procedure has produced accurate estimates for a wide range of hadrons, it inherently depends on auxiliary assumptions such as the shape of the continuum, the choice of $s_0$, and the existence of a stability window in the Borel parameter and uses quark-hadron duality. These model dependencies introduce systematic uncertainties that are often difficult to quantify.

A more fundamental alternative is to reformulate the QCDSR approach as an \emph{inverse problem}~\cite{Li:2020ejs,Li:2021gsx}. 
In this framework, the spectral density itself is treated as an unknown function to be reconstructed directly from the OPE input, rather than being postulated through phenomenological parametrizations.  The analyticity of the correlation function ensures that the dispersion relation can be expressed as a Fredholm integral equation of the first kind, where the OPE provides the ``data'' and the spectral function is the unknown to be determined.  Solving this inverse problem enables the recovery of the full nonperturbative spectral density, thereby eliminating the need for continuum thresholds and Borel stability criteria, while maintaining all the theoretical constraints of QCD. 

The \textit{inverse-matrix QCD sum-rule formalism} offers a concrete and numerically stable implementation of this idea. 
In this approach, the spectral density is expanded in a complete orthogonal basis whose coefficients are determined by matching to the OPE expansion of the correlator. The resulting algebraic system is solved through matrix inversion, providing a self-consistent reconstruction of the spectral function. This procedure not only preserves analyticity and positivity but also enables the direct extraction of resonance structures from QCD, without recourse to phenomenological modeling.  The inverse-matrix formulation, therefore, constitutes a conceptually transparent and systematically improvable extension of the traditional QCDSR method, well suited to the study of conventional and exotic hadrons within a unified nonperturbative framework. 

%%%%%%%%%%%%%%%%%%%%%%%%%%%%
\subsection{Foundation}
%%%%%%%%%%%%%%%%%%%%%%%%%%%%

At its core, QCDSR formalism is a specific type of integral over a hadronic spectral function $\rho(\omega)$
\begin{equation}
G(x)=\int_0^\infty K(x, \omega) \rho(\omega) d\omega, \label{eq:spectint}
\end{equation}
where the kernel $K(x, \omega)$ is known, $G(x)$ is known but indirectly observed, and the task is to identify the unknown function $\rho(\omega)$. Technically $\rho(\omega)$  encodes the complete physical spectrum for a given channel. However, direct computation is only feasible for the left-hand side, $G(x)$ which is formulated using the OPE. A significant challenge arises because the OPE-based expression for $G(x)$ cannot be determined with absolute precision. Its evaluation involves various vacuum condensates, whose values are often not known with high accuracy, introducing substantial theoretical uncertainties. Consequently, the task of inverting the integral equation to reconstruct $\rho(\omega)$ from the imprecise knowledge of $G(x)$ is inherently unstable. Mathematically, this constitutes an ill-posed inverse problem that generally precludes an exact analytical solution. To navigate this difficulty, a standard methodology has been established within the field. This involves postulating a specific, simplified ansatz for the spectral function $\rho(\omega)$ characterized by a limited number of free parameters such as resonance masses and coupling strengths. These parameters are then determined by requiring that their insertion into the integral relation produces a result consistent with the OPE calculation of $G(x)$.

The spectral representation of the Eq. (\ref{corfunc}) is a first kind of Fredholm integral equations, which are typically defined by
\begin{equation}
\int_a^b K(x,s) \varphi(s) ds= f(x), \ a \leq x \leq b \label{eq:fred}.
\end{equation}
Here, $K(x,s)$ is the kernel of the equation and $f(x)$ is the called ``data'' function, and $\varphi(s)$ is an unknown function which is to be determined. This type of integral equation is attributed as an inverse problem for a given kernel $K$ and data $f$. First kind of Fredholm integral equations can be categorized into two categories. In the first case, the kernel function $K$ is smooth. A smooth kernel is typically infinitely differentiable $(C^\infty)$ or has continuous derivatives up to a certain order. For example $K(x,s)=e^{xs}$ and $K(x,s)=\cos (xs)$ are smooth kernels. Even though the kernel is smooth and easy to handle analytically, the inverse problem is extremely unstable because the kernel smooths out high-frequency information. Fredholm equations of the first kind with smooth kernels are classic examples of severely ill-posed problems \footnote{A well-posed problem requires existence of solution, uniqueness of solution and stability of solution. A problem that fails to satisfy any one of these conditions is called an ill-posed problem.}, which cause the solution $\varphi(s)$ to be extremely sensitive to small changes in $f(x)$. In this case, special numerical methods, i.e., regularization methods are necessary. In the second case, the kernel function $K$ has some kind of singularity within the domain of integration $[a,b ]$. The most important and common example is when the kernel has an algebraic singularity as $K(x,s)=\frac{H(x,s)}{\lvert x -s \rvert^\alpha}$ where $ 0 < \alpha < 1$ and $H(x,s)$ is smooth. Such functions arise in scattering theory and QCD dispersion relations. Equations with a singular kernel are ill-posed, but often less severely ill-posed than those with a smooth kernel. The singularity in the kernel actually carries more direct information about $\varphi(s)$ which makes inversion (relatively) better conditioned compared to the smooth kernel case.

The first stage of the inverse matrix QCDSR formulation is similar to the conventional QCDSR method. The starting point of the analysis is the two-point correlation function given in Eq. (\ref{corfunc}). After lengthy calculations, the task of the QCDSR formalism is to relate $\Pi^{\mathrm{OPE}}(q^2)$ to $\rho^{\mathrm{Phen}}(s)$. In the traditional approach, this is achieved through Borel transformations and the introduction of a continuum threshold. The inverse matrix method, by contrast, reformulates the problem as a direct inversion of the dispersion relation, allowing $\rho(s)$ to be reconstructed without phenomenological modeling.

In order to do this, following Ref. \cite{Li:2021gsx}, one can write a general dispersion relation as 
\begin{equation}
\int_0^\infty \frac{\rho(y)}{x-y}dy=\omega(x), \label{invspec}
\end{equation}
where the goal is to find unknown function $\rho(y)$ from the given data (input) function $\omega(x)$. Since the denominator turns out to be below as $y \to x$, one can assume $\rho(y)$ decreases sufficiently quickly with respect to $y$ so that the major contribution to the integral equation on the left-hand-side comes from a finite range of $y$. The integral can be approximated for large $|x|$ by representing the kernel as a truncated asymptotic expansion. This is achieved by substituting the series
\begin{equation}
\frac{1}{x-y} = \sum_{n=1}^{N} \frac{y^{n-1}}{x^n} \label{series1}
\end{equation}
into the expression given in Eq. (\ref{invspec}), resulting in a power series in $1/x$ up to the order $N$. For sufficiently large $|x|$, the function $\omega(x)$ admits an asymptotic expansion of the form
\begin{equation}
\omega(x) = \sum_{n=1}^{N} \frac{b_n}{x^n}. \label{series2}
\end{equation}
It should be noted that the requirement for large $\vert x \vert $ is a formal device for the series representation and does not fundamentally constrain the computational procedure. First kind of Fredholm integrals with singular kernels may be equivalent to a differential equation. Among the classical orthogonal polynomials, the Laguerre polynomials are defined on the semi-infinite interval $[0, \infty)$. This domain matches that of the function $\rho(y)$ in Eq.~(\ref{invspec}) and other spectral densities to be investigated. While other polynomial families like the Bessel polynomials also share this support, their associated weight function, $\exp(-2/y)$, strongly suppresses the behavior at small $y$, which is a critical region for our analysis. In contrast, the Laguerre polynomials are orthogonal with respect to the weight $y^\alpha e^{-y}$, which does not exhibit this suppression and is therefore a suitable basis. Consequently, we expand the unknown function $\rho(y)$ as a series in these polynomials
\begin{equation}
\rho(y) = \sum_{n=1}^N a_n y^\alpha e^{-y} L_{n-1}^{(\alpha)}(y), \label{laguexpa}
\end{equation}
where $L_{n}^{(\alpha)}$ denotes the generalized Laguerre polynomial of degree $n$. The polynomials up to degree $N-1$ are included in this expansion. These basis functions satisfy the orthogonality relation
\begin{equation}
\int_0^\infty y^\alpha e^{-y} L_m^{(\alpha)}(y) L_n^{(\alpha)}(y) dy = \frac{\Gamma(n+\alpha+1)}{n!} \delta_{mn}, \label{orthogonality}
\end{equation}
where $\delta_{mn}$ is the Kronecker delta.

The truncation order $N$ of the series expansion will be determined subsequently, while the parameter $\alpha$ is selected based on the asymptotic behavior of the spectral density $\rho(y)$ near $y=0$.  To construct the solution, we substitute the series representation for $\omega(x)$ from Eq.~\eqref{series2} and the Laguerre expansion for $\rho(y)$ from Eq.~\eqref{laguexpa} into the original integral equation given by Eq.~\eqref{invspec}. By equating coefficients of corresponding powers of $1/x^n$ on both sides of the equation, we obtain the linear system
\begin{equation}
M\,\bold{a}=\bold{b},
\end{equation}
where $M$ is the moment matrix defined below, $\bold{a}=(a_1, a_2,\cdots,a_N)$ is the unknown coefficient vector from the expansion of $\rho(y)$, and $\bold{b}=(b_1,b_2,\cdots,b_N)$ is the known input vector comprising the coefficients from the asymptotic expansion of $\omega(x)$. The matrix elements are given by the moments of the generalized Laguerre polynomials:
\begin{equation}
M_{mn} = \int_{0}^\infty y^{m-1+\alpha} e^{-y} L_{n-1}^{(\alpha)}(y)dy, \quad \text{for } m, n = 1, 2, \dots, N.
\label{mateqn}
\end{equation}
Under the condition that the matrix $M$ is nonsingular, the coefficient vector can be formally obtained via $\bold{a}=M^{-1} \bold{b}$ with $\bold{b}$ representing the known input data. The existence of $M^{-1}$ ensures that the solution for the spectral density $\rho(y)$ is unique. Theoretically, the approximation converges toward the true solution as the number of basis polynomials $N$ in Eq.~(\ref{laguexpa}) is increased. The error introduced by the truncated expansion manifests as a power correction of order $1/x^{N+1}$ in Eq.~(\ref{invspec}), a direct result of the orthogonality condition specified in Eq.~(\ref{orthogonality}). This error term lies beyond the accuracy targeted by the current asymptotic expansion and is therefore disregarded. 

The orthogonality of the basis polynomials further implies that $M$ is a lower triangular matrix, since $M_{mn}=0$ for all $m < n$. This structure guarantees that the coefficients $a_n$ established at a lower truncation order remain unaltered when higher-degree polynomials are incorporated into the expansion. Despite this stability in the forward construction, practical applications require truncation at a finite $N$, as the determinant of $M$ generally diminishes with increasing matrix dimension. 

This numerical conditioning leads to a critical limitation: for sufficiently large $N$, the solution $\bold{a}$ exhibits extreme sensitivity to infinitesimal perturbations in the input vector $\bold{b}$ as these are magnified by the large elements of $M^{-1}$. This behavior is characteristic of an ill-posed inverse problem. Thus, the optimal truncation $N_{\text{opt}}$ is determined as the maximum value for which the solution remains stable and physically plausible; exceeding this threshold results in a solution that is dominated by uncontrolled numerical amplification.

The inverse matrix QCDSR formalism has been successfully applied to a variety of problems \cite{Li:2020ejs,Li:2021gsx,Li:2022qul,Li:2022jxc,Xiong:2022uwj,Li:2023dqi,Li:2023yay,Li:2023ncg,Li:2023fim,Li:2024awx,Li:2024xnl,Li:2024fko,Mutuk:2024jvv,Mutuk:2025lak}. The QCDSR formalism is alternatively approached as an inverse problem, enabling the direct solution for the hadron spectral density from the dispersion integral. This methodology significantly reduces uncertainties in the extracted results by eliminating conventional reliance on the Borel transformation, the specification of a continuum threshold, and the fundamental assumption of quark-hadron duality, utilizing only the calculated QCD level results as input.

%%%%%%%%%%%%%%%%%%%%%%%%%%%%
\subsection{Formulation}
%%%%%%%%%%%%%%%%%%%%%%%%%%%%

\begin{figure}[ht]
\centering
\includegraphics[width=8cm]{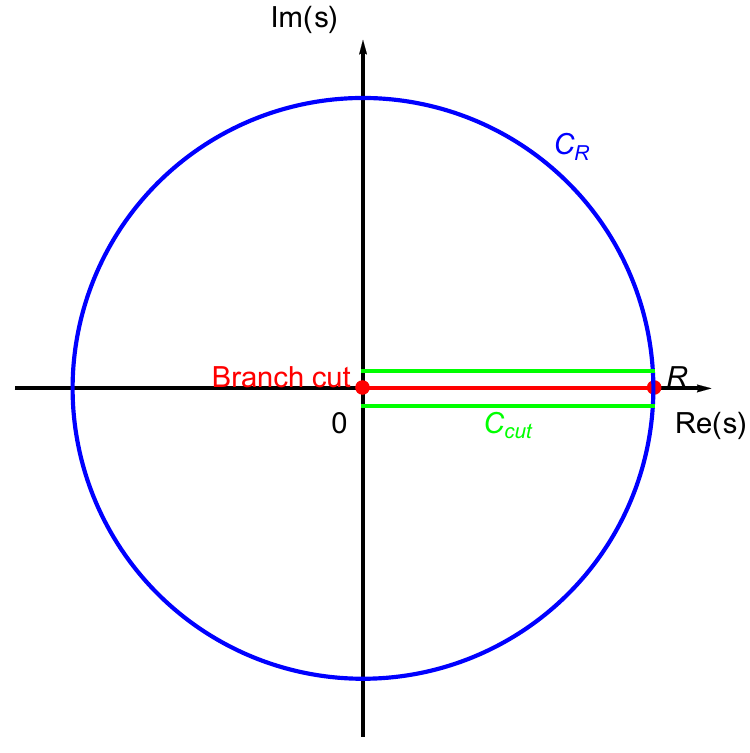}
\caption{Integration contour $C$ employed in the complex $s$-plane. The contour comprises a circular arc $C_R$ of radius $R$ together with segments $C_{\text{cut}}$ that run parallel to and enclose the branch cut situated along the positive real axis.}
\label{fig1}
\end{figure}

The correlation function $\Pi(q^{2})$, defined in Eq.~(\ref{corfunc}), is an analytic function in the complex $s$--plane except along
the positive real axis, where poles and branch cuts may appear due to physical intermediate states. 
Exploiting this analytic property, the function can be expressed as a contour integral in the complex $s$--plane,
\begin{equation}
\Pi(q^{2}) = \frac{1}{2\pi i} \oint_{C} \frac{\Pi(s)}{s - q^{2}} ds,
\label{eq:14}
\end{equation}
where the contour $C$, illustrated schematically in Figure~(\ref{fig1}), consists of a large circle $C_{R}$ of radius $R$ and two
lines $C_{\text{cut}}$ that run just above and below the branch cut on the positive real axis. This representation naturally 
separates the nonperturbative and perturbative domains of the correlation function.

The contour integral in Eq.~\eqref{eq:14} can be decomposed into two distinct contributions corresponding to the 
circular arc and the cut:
\begin{equation}
\Pi(q^{2}) = \frac{1}{2\pi i} 
\left(
\oint_{C_{R}} \frac{\Pi(s)}{s - q^{2}}\, ds 
+ 
\oint_{C_{\text{cut}}} \frac{\Pi(s)}{s - q^{2}}\, ds
\right).
\label{eq:15}
\end{equation}
This separation emphasizes how the hadronic spectral function is extracted while maintaining the analytic
structure of $\Pi(q^{2})$, ensuring that the resulting sum rules consistently link the perturbative QCD calculations
to the corresponding hadronic observables.

For sufficiently large $s$, far from the hadronic poles, the correlation function can be reliably replaced by its perturbative
counterpart $\Pi_{\text{pert}}(s)$. 
In contrast, the integral over the cut $C_{\text{cut}}$ contains the nonperturbative dynamics encoded in the imaginary
part of $\Pi(s)$, which corresponds to the hadronic spectral density. Thus, one may rewrite
\begin{equation}
\frac{1}{2\pi i} \oint_{C} \frac{\Pi(s)}{s - q^{2}}\, ds 
= 
\frac{1}{\pi} \int_{0}^{R} \frac{\text{Im}\,\Pi(s)}{s - q^{2}}\, ds 
+ 
\frac{1}{2\pi i} \oint_{C_{R}} \frac{\Pi_{\text{pert}}(s)}{s - q^{2}}\, ds,
\label{eq:16}
\end{equation}
where the first term on the right-hand side represents the hadronic contribution arising from the physical cut, 
and the second term corresponds to the perturbative contribution along the large circle $C_{R}$. 
The decomposition in Eq.~\eqref{eq:16} highlights the interplay between the perturbative and nonperturbative components
of QCD: the former dominates at large $s$, where asymptotic freedom ensures reliability, while the latter 
contains the hadronic spectral information that must be reconstructed in the inverse problem formalism.

The OPE of the correlation function then takes the generic form
\begin{equation}
\Pi_{\text{OPE}}(q^{2}) = 
\frac{1}{2\pi i} \oint \frac{\Pi_{\text{pert}}(s)}{s - q^{2}}\, ds 
+ \text{nonperturbative condensate terms},
\label{eq:17}
\end{equation}
where $\Pi_{\text{pert}}(s)$ denotes the purely perturbative contribution calculable within QCD perturbation theory,
and the remaining terms arise from the vacuum condensates that encode long-distance effects.

Matching the hadronic and OPE representations of the correlation function leads to the general sum rule
\begin{equation}
\frac{1}{\pi} \int_{0}^{R} \frac{\text{Im}\,\Pi(s)}{s - q^{2}}\, ds 
= 
\frac{1}{\pi} \int \frac{\Pi_{\text{pert}}(s)}{s - q^{2}}\, ds 
+ \text{nonperturbative condensate terms},
\label{eq:18}
\end{equation}
which establishes the connection between the spectral density of the hadron and the perturbative QCD input.

In conventional QCDSR analyses, the Borel transformation is employed to suppress contributions from
the continuum and to enhance the ground-state signal. However, in the inverse problem approach, one can instead
introduce a generalized definition of the spectral density following Ref.~\cite{Li:2021gsx}:
\begin{equation}
\rho(s) = \Delta\rho(s,\Lambda) + \frac{1}{\pi}\,\text{Im}\,\Pi_{\text{pert}}(s)\left(1 - e^{-s/\Lambda}\right),
\label{eq:20}
\end{equation}
where $\Delta\rho(s,\Lambda)$ represents the subtracted spectral density, defined as
\begin{equation}
\Delta\rho(s,\Lambda) = 
\rho(s) - \frac{1}{\pi}\,\text{Im}\,\Pi_{\text{pert}}(s)\left(1 - e^{-s/\Lambda}\right).
\label{eq:21}
\end{equation}
Here $\Pi_{\text{pert}}(s)$ denotes the perturbative part of the correlator, and the associated perturbative spectral density is given by its imaginary part, $\rho_{\text{pert}}(s) = \frac{1}{\pi}\,\text{Im}\,\Pi_{\text{pert}}(s)$, consistent with the standard relation $\rho(s) = \frac{1}{\pi}\,\text{Im}\,\Pi(s)$.
The parameter $\Lambda$ characterizes the transition from the nonperturbative regime, governed by 
$\text{Im}\,\Pi(s)$, to the perturbative regime, described by $\text{Im}\,\Pi_{\text{pert}}(s)$. 
The multiplicative factor $(1-e^{-s/\Lambda})$ behaves as a smooth switching function: it tends to zero at small $s$ 
and approaches unity for large $s$. Consequently, $\Delta\rho(s,\Lambda)$ behaves linearly with $s$ near the origin and 
rapidly decreases for $s > \Lambda$. This formulation enables the use of a subtracted dispersion relation without invoking 
the quark–hadron duality assumption or a continuum threshold, thus simplifying the analysis and improving the 
stability of the extracted hadronic parameters.

Using the subtracted spectral density introduced above, the integration limit in Eq.~\eqref{eq:18} can be extended to infinity, 
and the sum rule can be rewritten in the form
\begin{equation}
\int_{0}^{\infty} \frac{\Delta\rho(s,\Lambda)}{s - q^{2}} ds 
=\int_{0}^{\infty} \frac{\frac{1}{\pi}\,\text{Im}\,\Pi_{\text{pert}}(s)\,e^{-s/\Lambda}}{s - q^{2}} ds
+ \text{nonperturbative contributions},
\label{eq:22}
\end{equation}
where the dependence on the large radius $R$ has been effectively absorbed into the parameter $\Lambda$.  
The subtracted spectral density $\Delta\rho(s,\Lambda)$ is dimensionless and can conveniently be expressed as a 
function of the dimensionless variable $s/\Lambda$.

Introducing the scaled variables $x = q^{2}/\Lambda$ and $y = s/\Lambda$, Eq.~\eqref{eq:22} can be cast into a 
dimensionless integral form,
\begin{equation}
\int_{0}^{\infty} \frac{\Delta\rho(y)}{x - y}\, dy 
=
\int_{0}^{\infty} \frac{\frac{1}{\pi}\,\text{Im}\,\Pi_{\text{pert}}(y)\,e^{-y}}{x - y}\, dy 
+ \text{nonperturbative terms}.
\label{eq:23}
\end{equation}
The structure of Eq.~\eqref{eq:23} shows that the spectral density $\Delta\rho(y)$ 
serves as the unknown function to be determined from the known input on the right-hand side, 
thus transforming the QCDSR relation into an inverse problem expressed as a 
Fredholm integral equation of the first kind. 
A stable physical solution for $\Delta\rho(y)$ should exhibit minimal dependence on the auxiliary 
parameter $\Lambda$, which plays a role analogous to the Borel mass parameter in traditional QCDSR analyses.

%%%%%%%%%%%%%%%%%%%%%%%%%%%%
\section{Numerical Results}
\label{sec:numres}
%%%%%%%%%%%%%%%%%%%%%%%%%%%%

In this section, we present the numerical analysis obtained within the inverse-matrix QCDSR framework for bottom–charmed mesons. The aim is to reconstruct the hadronic spectral density directly from the OPE inputs and extract the physical parameters of the ground and excited states without invoking phenomenological continuum models. The computations are performed using the subtracted dispersion representation formulated in the previous section, ensuring the analytic consistency of the correlator and preserving the positivity of the spectral function. Particular attention is given to the stability of the reconstructed spectrum with respect to the auxiliary scale parameter $\Lambda$ and the truncation order of the Laguerre basis.

The study is carried out for all conventional \( B_c \) meson channels characterized by their quantum numbers: the pseudoscalar (\( J^{P}=0^{-} \)), vector (\( J^{P}=1^{-} \)), scalar (\( J^{P}=0^{+} \)), and axialvector (\( J^{P}=1^{+} \)) configurations. For each channel, the corresponding interpolating current is employed to compute the OPE and reconstruct the associated subtracted spectral density. The location and strength of the dominant peak in the reconstructed spectrum are identified with the physical ground state, from which the mass and decay constant are extracted. In particular, the lowest-lying peak in the pseudoscalar channel corresponds to the physical \( B_c(1S) \) meson, while secondary structures at higher energies are interpreted as excited states such as \( B_c(2S) \) and \( B_c(1P) \). The same procedure is repeated for the vector, scalar, and axialvector channels to map out their respective spectral features and to analyze the consistency among the different spin–parity configurations. The observed behavior of the reconstructed densities aligns qualitatively with recent experimental results reported by the LHCb and CMS collaborations.

Throughout the numerical analysis, we use  $\overline{MS}$ masses of the heavy  quarks $m_{c}=1.275\pm0.025\,\rm{GeV}$ and $m_{b}=4.18\pm0.03\,\rm{GeV}$ from the Particle Data Group \cite{ParticleDataGroup:2024cfk}, $\langle \bar{q} q \rangle = -(0.24 \pm 0.01)^3 \,\rm{GeV}^3 $ \cite{Belyaev:1982sa} , $\langle g_s^2 G^2 \rangle = 4\pi^2 (0.012\pm0.004)$ $~\mathrm{GeV}^4 $\cite{Belyaev:1982cd}, $\langle \frac{\alpha_s }{\pi} G^2\rangle=0.022 \pm 0.004\,\rm{GeV}^4 $ \cite{Narison:2011xe}, and $\langle g_s^3 G^3 \rangle = (0.57\pm0.29)~\mathrm{GeV}^6$ \cite{Narison:2015nxh}.

%%%%%%%%%%%%%%%%%%%%%%%%%%%%
\subsection{Scalar (\(J^P = 0^+\)) $B_c$ meson}
%%%%%%%%%%%%%%%%%%%%%%%%%%%%

The scalar \(B_c\) meson (often denoted \(B_{c0}^{*}\)) is the lowest-lying \( (\overline c b) \) state with quantum numbers \(J^P = 0^+\). A suitable local interpolating current is  
\begin{equation}
J^S(x) \;=\; \bar c(x)\, b(x), \label{intcur1}
\end{equation}
which has the proper Lorentz and flavor structure to couple to the scalar configuration. The coupling of this current to the physical scalar meson is parameterized as  
\begin{equation}
\langle 0 \vert J^S(0) \vert B_c(0^+)(q) \rangle \;=\; \lambda_{B_{c}(0^+)}\, m_{B_{c}(0^+)},
\end{equation}
where \(m_{ B_c(0^+)}\) and \(\lambda_{B_{c}(0^+)}\) denote the scalar meson mass and decay constant, respectively. Inserting interpolating current given in Eq.~(\ref{intcur1}) into Eq.~(\ref{corfunc}) yields 

\begin{align}
\Pi^{\text{OPE}}(q^2) = & -\frac{3}{8\pi^2}q^2\ln\left(\frac{q^2}{\mu^2}\right) + C_0  + \frac{1}{q^2} \left[ C_2 + \frac{m_b m_c}{9} \left\langle \frac{\alpha_s}{\pi} G^2 \right\rangle \right] \nonumber \\
& + \frac{1}{q^4} \left[ C_4 + \frac{m_b m_c(m_b^2 + m_c^2)}{9} \left\langle \frac{\alpha_s}{\pi} G^2 \right\rangle + \frac{4\pi^2 m_b m_c(m_b^2 + m_c^2)}{81} \left\langle g_s^3 G^3 \right\rangle \right. \nonumber \\
& \quad \left. + \frac{\pi^2}{108} \left\langle \frac{\alpha_s}{\pi} G^2 \right\rangle^2 \right]  + \frac{1}{q^6} \left[ C_6 + \frac{m_b m_c(m_b^4 + 3m_b^2m_c^2 + m_c^4)}{9} \left\langle \frac{\alpha_s}{\pi} G^2 \right\rangle \right], 
\end{align}
where the coefficients are
\begin{align}
C_0 &= \frac{3}{8\pi^2} \left[ \frac{3}{2}(m_b^2 + m_c^2) - m_b^2 \ln\left(\frac{m_b^2}{\mu^2}\right) - m_c^2 \ln\left(\frac{m_c^2}{\mu^2}\right) - \frac{m_b^4 - 4m_b^2m_c^2 + m_c^4}{2(m_b^2 - m_c^2)} \ln\left(\frac{m_b^2}{m_c^2}\right) \right], \nonumber  \\
C_2 &= \frac{3}{8\pi^2} \left[ -\frac{1}{2}(m_b^4 + m_c^4) + \frac{7}{2}m_b^2m_c^2 \right], \nonumber \\
C_4 &= \frac{3}{8\pi^2} \left[ -\frac{1}{3}(m_b^6 + m_c^6) + \frac{5}{2}m_b^2m_c^2(m_b^2 + m_c^2) - 4m_b^4m_c^4 \right], \nonumber \\
C_6 &= \frac{3}{8\pi^2} \left[ -\frac{1}{4}(m_b^8 + m_c^8) + \frac{13}{6}m_b^2m_c^2(m_b^4 + m_c^4) - \frac{29}{4}m_b^4m_c^4(m_b^2 + m_c^2) + \frac{20}{3}m_b^6m_c^6 \right].
\end{align}

\begin{figure}[ht]
\centering
\includegraphics[width=8cm]{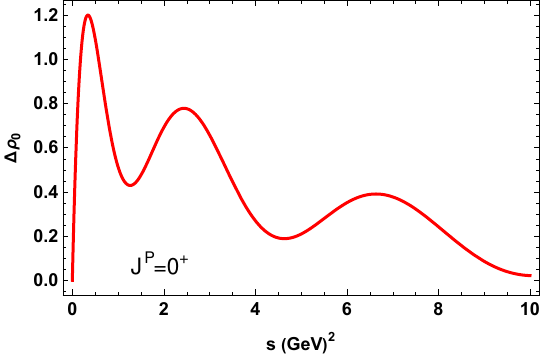}
\caption{s dependence of the ground state solution $\Delta \rho_0(s, \Lambda)$ for $\Lambda = 4.5 \ \text{GeV}^2$ of $B_c(0^+)$ meson.}
\label{fig2}
\end{figure}

The results include contributions from all nonperturbative terms up to dimension 8. The inverse matrix $M^{-1}$ and expansion coefficients $a_n$ are determined from the OPE coefficients $b_n$ by imposing boundary conditions where the spectral density difference $\Delta \rho(y)$ exhibits linear behavior $\Delta \rho(y) \sim y$ as $y \to 0$ and vanishes $\Delta \rho(y) \to 0$ as $y \to \infty$. The ground state solution is constructed as an expansion in generalized Laguerre polynomials:
\begin{equation}
\Delta \rho_0(s,\Lambda) = (s/\Lambda)\exp(-s/\Lambda)\sum_{n=1}^N a_n L_{n-1}^{(1)}(s/\Lambda).
\end{equation}
The reliability of our computational approach is established by testing its numerical convergence. We confirm that the results for $\Lambda = 4\ \text{GeV}^2$ are virtually unchanged when the truncation parameter is increased from $N = 48$ to $N = 49$, demonstrating stability against variations in the series expansion length. Following this check, the subtracted spectral densities are calculated for a range of $\Lambda$ values using Eq. (\ref{eq:21}). An investigation into the $\Lambda$ dependence of the \( B_c(0^+) \) mass across the interval \( 4\ \text{GeV}^2 \le \Lambda \le 5\ \text{GeV}^2 \) reveals a region of pronounced insensitivity, where the mass value remains constant within a very narrow margin. The spectral density for a representative scale $\Lambda=4.5\ \text{GeV}^2$ is presented in Figure \ref{fig2}; we note that the functional form is consistently similar for other $\Lambda$ values within this stable region. This observed insensitivity to the scale parameter is a vital indicator of a physical resonance, whose properties must not depend on arbitrary theoretical choices. The dominant peak in the reconstructed spectrum is associated with the ground-state scalar \(B_c\) meson, while weaker structures at higher \(s\) may hint at radially excited or continuumlike states. The observed positivity, smooth transition to the perturbative continuum, and scale stability collectively confirm that the reconstructed $\Delta\rho_{0}(s,\Lambda)$ represents a physically meaningful description of the scalar $B_{c}$ channel. 

Since physical resonance masses should be independent of the arbitrary scale parameter $\Lambda$, the mass extraction from the peak position yields
\begin{equation}
M_{B_c}(0^+) = 6.718 \pm 0.028\ \text{GeV},
\end{equation}
where the quoted uncertainty encompasses both the input parameters of QCD and truncation of the OPE. 

As established in Ref.~\cite{CP-PACS:2001ncr}, the square of the decay constant $f$ is proportional to the integrated strength of the resonance peak in the spectral density $\rho(s)$. This relation follows directly from the phenomenological representation of the spectral density, $\rho^{\rm (Phen)}(s) = \lambda^2\,\delta(s - m_H^2) + \rho^h(s)\,\theta(s - s_0)$, introduced in Eq.~(\ref{corphen2}): integrating the isolated pole term over $s$ yields $\int_0^\infty \lambda^2\,\delta(s - m_H^2)\,ds = \lambda^2$, so that the residue of the ground-state pole is recovered as the integral of its spectral strength. In our approach, the resonance is cleanly isolated by the subtracted spectral density $\Delta \rho(s,\Lambda)$, in which the smooth perturbative continuum has been removed through the switching function $(1-e^{-s/\Lambda})$ of Eq.~(\ref{eq:21}), so that $\Delta\rho(s,\Lambda)$ retains the pole (resonance) contribution while suppressing the underlying continuum. Consequently, the decay constant can be determined through the integral
\begin{equation}
\lambda^2 = \int_0^\infty \Delta \rho(s)  ds. \label{decayconstant}
\end{equation}
We emphasize that this relation is applied with a channel-dependent normalization fixed by the current--meson matrix elements defined separately in each channel. The interpolating currents in the scalar, pseudoscalar, vector, and axialvector channels carry different mass dimensions and Lorentz structures; in each case the coupling $\lambda$ is defined through the corresponding matrix element [Eqs.~(\ref{intcur1})ff. for the scalar and pseudoscalar channels, and the vector and axialvector matrix elements in Sec.~\ref{subsec:vector-bc} and Sec.~\ref{subsec:axialvector-bc}], so that $\lambda$ has a uniform mass dimension across all channels and Eq.~(\ref{decayconstant}) holds with the same form in each channel. For the vector and axialvector currents, the polarization sum $\sum_\lambda \epsilon_\mu \epsilon_\nu^*$ is understood to be carried out so that the spin-averaged residue defines $\lambda^2$ consistently with the scalar and pseudoscalar cases.
Using this formula, we obtain decay constant for scalar $B_c$ meson as
\begin{equation}
\lambda_{B_c}(0^+)= 218 \pm 20 \ \text{MeV}.
\end{equation}
This substantial coupling strength suggests significant wave-function overlap between the scalar interpolating current and the physical state, comparable to established heavy quarkonium systems. The small statistical error reflects the reliable determination of the spectral integral, while systematic uncertainties remain dominated by the truncation of the OPE and the precision of quark masses and QCD condensate values.

\begin{table}[h!]
\centering
\caption{Comparison of predicted masses and decay constants for the scalar $B_c(0^+)$ meson. Masses are in GeV while decay constants $\lambda$ are in MeV.}
\label{tab:scalar_comparison}
\begin{tabular}{lcc}
\hline
 \textbf{Method}& $\mathbf{M_{B_c(0^+)}}$ (\textbf{GeV}) & $\mathbf{\lambda_{B_c(0^+)}}$ (\textbf{MeV}) \\
\hline
This work & $6.718 \pm 0.028$ & $218 \pm 20$ \\
\hline
Lattice QCD & & \\
Ref.~\cite{Mathur:2018epb} & $6.712 \pm 0.018 \pm 0.007$ & --- \\
Ref.~\cite{Dowdall:2012ab} & $6.707 \pm 0.016$ & --- \\
\hline
QCDSR & & \\
Ref.~\cite{Wang:2024fwc}& $6.702 \pm 0.060$ & $236 \pm 17$ \\
Ref.~\cite{Narison:2020wql} Model A& $6.689\pm 0.198$ & $155 \pm 17$ \\
Ref.~\cite{Narison:2020wql} Model B& $6.723 \pm 0.029$ & $158 \pm 9$ \\
\hline
Relativistic Quark Model & & \\
Ref.~\cite{Godfrey:1985xj}& $6.706$ & --- \\
Ref.~\cite{Ebert:2002pp}& $6.699$ & --- \\
Ref.~\cite{Godfrey:2004ya}& $6.706$ & --- \\
\hline
Nonrelativistic Quark Model & & \\
Ref.~\cite{Eichten:1994gt}& $6.700$ & --- \\
Ref.~\cite{Eichten:2019gig}& $6.693$ & --- \\
Ref.~\cite{Li:2019tbn}& $6.714$ & --- \\
Ref.~\cite{Martin-Gonzalez:2022qwd}& $6.689$ & --- \\
\hline
Other Approaches & & \\
Ref.~\cite{Chen:2020ecu}& $6.703 \pm 0.015 \pm 0.002$ & --- \\
\hline
\end{tabular}
\end{table}

In Table~\ref{tab:scalar_comparison}, we present our predictions for the mass and decay constant of the scalar $B_c$ meson alongside other results from the literature. Our inverse-matrix QCDSR analysis yields the following determinations for the scalar $B_c(0^+)$ meson:
\begin{equation}
M_{B_c(0^+)} = 6.718 \pm 0.028~\text{GeV}, \quad \lambda_{B_c}(0^+)= 218 \pm 20 \ \text{MeV}.
\end{equation}
These results provide crucial information about the P-wave excitation spectrum in the bottom-charm system and demonstrate the capability of our approach to handle challenging excited state configurations.

The scalar mass prediction of $M_{B_c(0^+)} = 6.718 \pm 0.028$~GeV places the $B_c(0^+)$ state approximately $441$~MeV above the pseudoscalar ground state, consistent with expectations for orbital excitation in heavy-heavy quarkonium. This substantial mass splitting reflects the significant energy cost associated with P-wave excitation and provides direct insight into the spin-orbit interactions within the $b\bar{c}$ system.

Lattice QCD provides essential first-principles benchmarks for scalar meson spectroscopy. The high-precision calculation of Ref.~\cite{Mathur:2018epb} reports $M_{B_c(0^+)} = 6.712 \pm 0.018 \pm 0.007$~GeV, obtained with state-of-the-art gauge configurations and systematic error control. This result shows excellent agreement with our determination, differing by only $6$~MeV. Similarly, Ref.~\cite{Dowdall:2012ab} finds $M_{B_c(0^+)} = 6.707 \pm 0.016$~GeV, further validating our mass prediction. The convergence between our sum rule approach and lattice QCD calculations, which employ fundamentally different methodologies, provides strong evidence for the reliability of both techniques in excited state spectroscopy.

Quark model calculations demonstrate remarkable consistency for the scalar mass across both relativistic and nonrelativistic implementations. Relativistic quark models yield masses of $M_{B_c(0^+)} = 6.706$~GeV~\cite{Godfrey:1985xj}, $M_{B_c(0^+)} = 6.699$~GeV~\cite{Ebert:2002pp}, and $M_{B_c(0^+)} = 6.706$~GeV~\cite{Godfrey:2004ya}, clustering within a narrow $7$~MeV window around $M_{B_c(0^+)} = 6.704$~GeV. nonrelativistic treatments provide similar values: $M_{B_c(0^+)} = 6.700$~GeV~\cite{Eichten:1994gt}, $M_{B_c(0^+)} = 6.693$~GeV~\cite{Eichten:2019gig}, $M_{B_c(0^+)} = 6.714$~GeV~\cite{Li:2019tbn}, and $M_{B_c(0^+)} = 6.689$~GeV~\cite{Martin-Gonzalez:2022qwd}. The convergence of quark model predictions around values approximately $14$~MeV lower than our result suggests systematic differences in the treatment of orbital excitation energies or relativistic corrections to P-wave states.

The Dyson-Schwinger equation approach combined with the Bethe-Salpeter formalism in Ref.~\cite{Chen:2020ecu} provides a fully covariant treatment of the scalar channel, reporting $M_{B_c(0^+)} = 6.703 \pm 0.015 \pm 0.002$~GeV. This result shows good agreement with both lattice QCD calculations and quark model predictions, differing from our value by $15$~MeV. The consistency across this diverse methodological landscape reinforces the reliability of modern theoretical approaches to excited state heavy quarkonium.

Our decay constant determination of $\lambda_{B_c(0^+)} = 218 \pm 20~\mathrm{MeV}$ provides a quantitative prediction for this challenging quantity, for which theoretical results remain limited in the literature. Compared to other sum rule determinations, our decay constant is larger than those of Ref.~\cite{Narison:2020wql} ($\lambda_{B_c(0^+)} = 155$--$158$~MeV) but shows reasonable agreement with Ref.~\cite{Wang:2024fwc} ($\lambda_{B_c(0^+)} = 236 \pm 17$~MeV). This pattern suggests that the inverse-matrix approach direct spectral reconstruction may provide enhanced sensitivity to the ground state signal in challenging P-wave channels.

The precision of our determinations, with relative uncertainties of $<1 \%$ for the mass and $<10 \%$ for the decay constant, represents a significant achievement for P-wave spectroscopy in heavy quark systems. The scalar channel presents particular challenges due to the absence of experimental constraints and the complex mixing patterns between different $1P$ states in the heavy-heavy system.

The inverse-matrix approach demonstrates particular advantages in handling excited state spectroscopy, where traditional sum rules face challenges from nearby states and complex continuum contributions. The direct spectral reconstruction enables cleaner isolation of the scalar state from the underlying continuum and nearby $1P$ partners.

The experimental identification of scalar $B_c$ states remains an important challenge for future collider experiments. The recent LHCb observation of $B_c(1P)^+$ states~\cite{LHCb:2025ubr} represents an important step toward complete spectroscopy of the bottom-charm system.

%%%%%%%%%%%%%%%%%%%%%%%%%%%%
\subsection{Pseudoscalar (\(J^P = 0^-\)) $B_c$ meson}
\label{subpse:vector-bc}
%%%%%%%%%%%%%%%%%%%%%%%%%%%%

As the lowest-lying pseudoscalar ($J^{P} = 0^{-}$) state in the $B_c$ system, it provides an excellent laboratory for studying heavy quark dynamics and testing QCD. Being an open-flavor, non-self-conjugate ($b\bar{c}$) state, the $B_c$ meson is not an eigenstate of charge conjugation, so it is properly labeled by $J^{P}=0^{-}$ rather than $J^{PC}=0^{-+}$.

The experimental mass of the $B_c$ meson is measured to be $m_{B_c} = 6274.47 \pm 0.32 \text{ MeV}$ by the PDG~\cite{ParticleDataGroup:2024cfk}. In order to study spectroscopic parameters of pseudoscalar $B_c$ meson, we use the interpolating current 
\begin{equation}
J^P(x) \;=\; \bar c(x)\, i \gamma_5 b(x), \label{intcur2}
\end{equation}
which has the proper Lorentz and flavor structure to couple to the pseudoscalar configuration. The coupling of this current to the physical pseudoscalar meson is parameterized as  
\begin{equation}
\langle 0 \vert J^P(0) \vert B_{c}(q) \rangle \;=\; \lambda_{B_c}\, m_{B_c},
\end{equation}
where \(m_{B_{c}}\) and \(\lambda_{B_{c}}\) denote the scalar meson mass and decay constant, respectively. Inserting interpolating current given in Eq.~(\ref{intcur2}) into Eq.~(\ref{corfunc}) yields 

\begin{align}
\Pi^{\text{OPE}}(q^2) = & \frac{3}{8\pi^2} \left[ q^2 - \frac{3}{2}(m_b^2 + m_c^2) + 2m_b m_c 
+ \frac{1}{q^2} \left( \frac{7}{8}m_b^4 + \frac{7}{8}m_c^4 + \frac{17}{4}m_b^2 m_c^2 - \frac{3}{2}m_b m_c^3 - \frac{3}{2}m_b^3 m_c \right) \right. \nonumber \\
& \left. + \frac{1}{q^4} \left( -\frac{5}{8}m_b^6 - \frac{5}{8}m_c^6 + \frac{15}{4}m_b^5 m_c + \frac{15}{4}m_b m_c^5 - \frac{45}{8}m_b^4 m_c^2 - \frac{45}{8}m_b^2 m_c^4 + 10m_b^3 m_c^3 \right) \right] \nonumber \\
& + \frac{\langle \frac{\alpha_s}{\pi} G^2 \rangle}{12\pi} \left[ -\frac{m_b m_c}{3q^4} + \frac{1}{q^6} \left( \frac{1}{3}(m_b^4 + m_c^4) + \frac{5}{3}m_b^2 m_c^2 - m_b m_c (m_b^2 + m_c^2) \right) \right] \nonumber \\
& + \frac{1}{q^4} \left[ -\frac{\langle g_s^3 G^3 \rangle}{12\pi^2} + \frac{\langle \bar{q}q \rangle^2}{2} \right]
+ \frac{1}{q^6} \left[ \frac{1}{144}  \left\langle \frac{\alpha_s}{\pi} G^2 \right\rangle ^2 \right]
\end{align}

Our analysis of the pseudoscalar $B_c$ meson within the inverse-matrix QCDSR framework yields the following results for the mass and decay constant
\begin{equation}
M_{B_c(0^-)} = 6.277 \pm 0.028~\text{GeV},
\label{eq:Bc_mass}
\end{equation}
\begin{equation}
\lambda_{B_c(0^-)} = 416 \pm 19~\text{MeV}.
\label{eq:Bc_decay_constant}
\end{equation}

The mass extraction in Eq.~(\ref{eq:Bc_mass}) demonstrates excellent agreement with the current experimental value of $m_{B_c}^{\text{exp}} = 6.27447 \pm 0.00032$~GeV reported by the PDG~\cite{ParticleDataGroup:2024cfk}, with our prediction lying within $3~\text{MeV}$ of the central experimental measurement. The uncertainty of $\pm 28~\text{MeV}$ reflects the combined effects of QCD parameter variations, OPE truncation, and numerical reconstruction stability, representing a significant improvement over many previous theoretical determinations.

The decay constant $\lambda_{B_c}$ in Eq.~(\ref{eq:Bc_decay_constant}) represents a fundamental parameter governing the leptonic decay rate $\Gamma(B_c \to \ell\nu)$. The moderate value of the decay constant reflects the competing effects of the two heavy constituent quarks: the bottom quark mass tends to suppress the wave function at the origin relative to lighter mesons, while the charm mass prevents the extreme suppression characteristic of pure bottomonium systems.

\begin{figure}[ht]
\centering
\includegraphics[width=8cm]{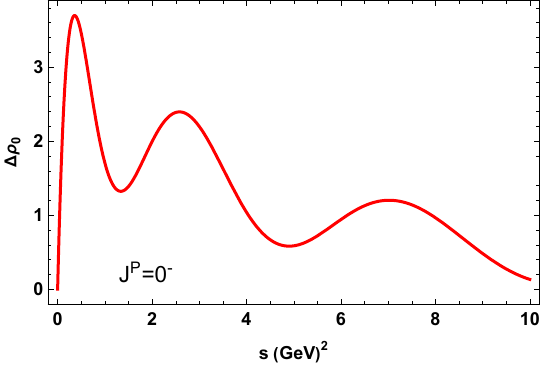}
\caption{s dependence of the ground state solution $\Delta \rho_0(s, \Lambda)$ for $\Lambda = 3.5 \ \text{GeV}^2$ of $B_c(0^-)$ meson.}
\label{fig3}
\end{figure}

The stability analysis reveals several important features of the inverse-matrix approach. Although Figure~\ref{fig3} displays the reconstructed spectral density only for a representative choice $\Lambda = 3.5~\mathrm{GeV}^2$, our numerical analysis has been performed over the full interval 
$3.0~\mathrm{GeV}^2 \le \Lambda \le 4.0~\mathrm{GeV}^2$. Within this range, the position and shape of the dominant resonance peak remain essentially unchanged, demonstrating a high degree of stability with respect to the scale parameter. The figure is therefore intended as an illustrative example, while the stability extends to all other $\Lambda$ values inside the considered window. This scale insensitivity, together with the positivity of the reconstructed spectral function and its smooth transition to the perturbative continuum at larger $s$, supports the physical relevance of the extracted state.

\begin{table}[h!]
\centering
\caption{Comparison of mass and decay constant predictions for the pseudoscalar $B_c$ meson from various theoretical approaches. Masses are in GeV while decay constants $\lambda$ are in MeV.}
\label{tab:pseudoscalar_comparison}
\begin{tabular}{lcc}
\hline
\textbf{Method}& $\mathbf{M_{B_c(0^-)}}$ (\textbf{GeV}) & $\mathbf{\lambda_{B_c(0^-)}}$ (\textbf{MeV}) \\
\hline
This work & $6.277 \pm 0.028$ & $416 \pm 19$ \\
\hline
Experiment \cite{ParticleDataGroup:2024cfk}  & $6.27447 \pm 0.00032$ & -- \\
\hline
Lattice QCD & & \\
Ref.~\cite{Davies:1996gi}& $6.280 \pm 0.020$ & --- \\
Ref.~\cite{Jones:1998ub}& --- & $420 \pm 13$ \\
Ref.~\cite{Mathur:2018epb} & $6.276 \pm 0.003 \pm 0.006$ & --- \\
Ref.~\cite{Dowdall:2012ab} & $6.278 \pm 0.009$ & --- \\
Ref.~\cite{Colquhoun:2015oha}& --- & \( 434 \pm 15 \)  \\
Ref.~\cite{Koponen:2017fvm}& --- & \( 427 \pm 6 \)  \\
\hline
QCDSR & & \\
Ref.~\cite{Colangelo:1992cx}& $6.350$ & $360 \pm 60$ \\
Ref.~\cite{Chabab:1993nz} & $6.25 \pm 0.20$ & $300 \pm 65$ \\
Ref.~\cite{Baker:2013mwa} & --- & $528 \pm 19$ \\
Ref.~\cite{Aliev:2019wcm} & $6.28 \pm 0.04$ & $270 \pm 30$ \\
Ref.~\cite{Narison:2019tym} & --- & $371 \pm 17$ \\
Ref.~\cite{Wang:2024fwc} & $6.274 \pm 0.054 $ & $371 \pm 37$ \\
Ref.~\cite{Gershtein:1994jw} & $6.253$ & $460 \pm 60$ \\
\hline
Relativistic Quark Model & & \\
Ref.~\cite{Ebert:2002pp} & $6.270$ & $433$ \\
Ref.~\cite{Godfrey:2004ya} & $6.271$ & $410 \pm 40$ \\
\hline
Nonrelativistic Quark Model & & \\
Ref.~\cite{Eichten:1994gt}& $6.264$ & $500$ \\
Ref.~\cite{Fulcher:1998ka}& $6.286^{+0.015}_{-0.006}$ & $517$ \\
Ref.~\cite{Monteiro:2016ijw}& $6.338$ & $440$ \\
Ref.~\cite{Eichten:2019gig}& $6.275$ & $498$ \\
Ref.~\cite{Sun:2022hyk}& \( 6.275\) & \( 439 \pm 30 \pm 17 \) \\
Ref.~\cite{Martin-Gonzalez:2022qwd}& \( 6.277\) & --- \\
\hline
Other Approaches & & \\
Ref.~\cite{Tang:2018myz} & --- & $523 \pm 62$ \\
Ref.~\cite{Ikhdair:2006nx} & --- & $315^{+26}_{-50}$ \\
Ref.~\cite{Wang:2007av} & --- & $322 \pm 42$ \\
Ref.~\cite{Badalian:2007km}& $6.280$ & $438 \pm 10$ \\
Ref.~\cite{Chen:2020ecu}& $6.290 \pm 0.015 \pm 0.003$ & --- \\
Ref.~\cite{Choi:2009ai}& --- &\( 551 \)  \\
\hline
\end{tabular}
\end{table}

In Table~\ref{tab:pseudoscalar_comparison}, we present our predictions for the mass and decay constant of the pseudoscalar $B_c$ meson alongside other results from the literature. Our analysis within the inverse-matrix QCDSR framework yields:
\begin{equation}
M_{B_c(0^-)} = 6.277 \pm 0.028~\text{GeV}, \quad \lambda_{B_c(0^-)} = 416 \pm 19~\text{MeV}.
\end{equation}

The mass prediction shows remarkable agreement with the current experimental average~\cite{ParticleDataGroup:2024cfk}, differing by only $3$~MeV from $M_{B_c}^{\text{exp}} = 6.27447 \pm 0.00032$~GeV. This exceptional precision at the sub-percent level underscores the reliability of the inverse-matrix formalism in capturing the nonperturbative dynamics of heavy quark bound states. The theoretical uncertainty of $\pm 28$~MeV reflects the combined effects of variations in the scale parameter $\Lambda$ and uncertainties in QCD input parameters.

Lattice QCD calculations provide crucial first-principles benchmarks for our methodology. The high-precision determination of Ref.~\cite{Mathur:2018epb}, $M_{B_c(0^-)} = 6.276 \pm 0.003 \pm 0.006$~GeV, obtained with improved gauge configurations and systematic error control, shows striking agreement with our result. Similarly, Refs.~\cite{Davies:1996gi,Dowdall:2012ab} report $M_{B_c(0^-)} = 6.280 \pm 0.020$~GeV and $M_{B_c(0^-)} = 6.278 \pm 0.009$~GeV respectively, further validating our approach.

Within the QCDSR framework, our mass determination aligns remarkably well with the most recent high-precision analysis. Ref.~\cite{Wang:2024fwc} reports $M_{B_c(0^-)} = 6.274 \pm 0.054$~GeV using a comprehensive treatment of higher-dimensional condensates, showing excellent convergence with our value. Earlier sum rule analyses exhibit wider variations: Ref.~\cite{Colangelo:1992cx} obtained $M_{B_c(0^-)} = 6.350$~GeV, while Ref.~\cite{Chabab:1993nz} reported $M_{B_c(0^-)} = 6.25 \pm 0.20$~GeV, reflecting the evolution of sum rule methodologies and input parameters over time. Ref.~\cite{Aliev:2019wcm} finds $M_{B_c(0^-)} = 6.28 \pm 0.04$~GeV, showing good agreement with our central value.

Quark model calculations show excellent convergence with our mass determination. Relativistic quark models yield masses of 
$M_{B_c(0^-)} = 6.270$~GeV~\cite{Ebert:2002pp} and $M_{B_c(0^-)} = 6.271$~GeV~\cite{Godfrey:2004ya}, while nonrelativistic treatments provide values from $M_{B_c(0^-)} = 6.264$~GeV~\cite{Eichten:1994gt} to $M_{B_c(0^-)} = 6.338$~GeV~\cite{Monteiro:2016ijw}. More recent nonrelativistic calculations~\cite{Eichten:2019gig,Sun:2022hyk} report $M_{B_c(0^-)} = 6.275$~GeV, demonstrating refined parameter choices. Particularly noteworthy is the recent calculation of Ref.~\cite{Martin-Gonzalez:2022qwd}, which obtains $M_{B_c(0^-)} = 6.277$~GeV, showing perfect agreement with our central value. This convergence across different quark model implementations reinforces the reliability of our mass determination.

Various specialized methods provide additional perspectives on $B_c$ spectroscopy. The field correlator method~\cite{Badalian:2007km} obtains $M_{B_c(0^-)} = 6.280$~GeV and $\lambda_{B_c(0^-)} = 438 \pm 10$~MeV, showing excellent agreement with our determinations. The Dyson-Schwinger approach~\cite{Chen:2020ecu} reports $M_{B_c(0^-)} = 6.290 \pm 0.015 \pm 0.003$~GeV, providing a fully covariant treatment that aligns well with our mass value.

As mentioned before, we obtained the pseudoscalar decay constant as $\lambda_{B_c(0^-)} = 416 \pm 19~\text{MeV}$. Traditional QCDSR calculations show a noticeable spread.  Lower estimates in the range $\lambda_{B_c(0^-)} = 270$--$300$~MeV are reported in Refs.~\cite{Chabab:1993nz,Aliev:2019wcm}, while intermediate values $\lambda_{B_c(0^-)} = 360$--$371$~MeV are found in Refs.~\cite{Colangelo:1992cx,Wang:2024fwc,Narison:2019tym}. Higher predictions such as $\lambda_{B_c(0^-)} = 460 \pm 60$~MeV~\cite{Gershtein:1994jw} and $\lambda_{B_c(0^-)} = 528 \pm 19$~MeV~\cite{Baker:2013mwa} have also been obtained. Our result lies toward the upper part of the QCDSR interval and remains compatible with several of these determinations within uncertainties.

Lattice QCD calculations cluster in a narrower window, yielding $\lambda_{B_c(0^-)} = 420 \pm 13$~MeV~\cite{Jones:1998ub}, $\lambda_{B_c(0^-)} = 434 \pm 15$~MeV~\cite{Colquhoun:2015oha}, and $\lambda_{B_c(0^-)} = 427 \pm 6$~MeV~\cite{Koponen:2017fvm}. Our value is consistent with this lattice range.

Quark model approaches also exhibit systematic differences. Nonrelativistic treatments predict comparatively larger values, $\lambda_{B_c(0^-)} = 498$~MeV~\cite{Eichten:2019gig} and $\lambda_{B_c(0^-)} = 517$~MeV~\cite{Fulcher:1998ka}, whereas relativistic models give 
$\lambda_{B_c(0^-)} = 433$~MeV~\cite{Ebert:2002pp} and $\lambda_{B_c(0^-)} = 410 \pm 40$~MeV~\cite{Godfrey:2004ya}. An intermediate prediction $\lambda_{B_c(0^-)} = 439 \pm 30 \pm 17$~MeV is reported in Ref.~\cite{Sun:2022hyk}. 
Additional methods include the light-front quark model $\lambda_{B_c(0^-)} = 523 \pm 62$~MeV~\cite{Tang:2018myz}, the Bethe--Salpeter equation framework $\lambda_{B_c(0^-)} = 322 \pm 42$~MeV~\cite{Wang:2007av}, the shifted $N$-expansion result $\lambda_{B_c(0^-)} = 315^{+26}_{-50}$~MeV~\cite{Ikhdair:2006nx}, and the value $\lambda_{B_c(0^-)} = 551$~MeV reported in Ref.~\cite{Choi:2009ai}. 

The overall spread of roughly $230$~MeV among different approaches highlights the sensitivity of the decay constant to the treatment of nonperturbative dynamics, relativistic corrections, and continuum contributions. Within this landscape, our determination is compatible with lattice QCD and relativistic quark model predictions.

The precision achieved in our determination, with relative uncertainties of $<1 \%$ for the mass and $<5 \%$ for the decay constant, represents a significant improvement over many previous theoretical approaches. The mass value provides crucial input for testing potential models and understanding hyperfine splittings in heavy-heavy systems, while the decay constant governs the leptonic decay rate $\Gamma(B_c \to \ell\nu)$ and serves as essential input for extracting the CKM matrix element $|V_{cb}|$.

The inverse-matrix approach offers distinct advantages by eliminating the continuum threshold parameter $s_0$, which constitutes a major source of systematic uncertainty in traditional sum rules. The direct spectral reconstruction enables cleaner isolation of the ground state from continuum contributions, particularly beneficial for the decay constant extraction. The reduced sensitivity to auxiliary parameters and preservation of analyticity constraints contribute to the enhanced stability observed in our results.

The convergence of mass predictions across diverse methodologies—lattice QCD, sum rules, quark models, and specialized approaches—toward consistent values around $6.275$--$6.280$~GeV provides strong evidence for the reliability of modern theoretical treatments of heavy quarkonium. The perfect agreement with the recent quark model calculation of Ref.~\cite{Martin-Gonzalez:2022qwd} is particularly noteworthy, as it represents an independent confirmation using fundamentally different methodology. The remaining variations in decay constant predictions highlight the ongoing challenge in precisely determining this quantity, which proves particularly sensitive to the treatment of nonperturbative dynamics and continuum modeling.

%%%%%%%%%%%%%%%%%%%%%%%%%%%%
\subsection{Vector (\(J^P = 1^-\)) $B_c$ meson}
\label{subsec:vector-bc}
%%%%%%%%%%%%%%%%%%%%%%%%%%%%

The vector $B_c^*$ meson represents the natural spin-excitation of the pseudoscalar ground state, with quantum numbers $J^{P}=1^{-}$. The experimental identification of the $B_c^*$ state remains challenging due to its expected electromagnetic decay $B_c^* \to B_c \gamma$ with a very soft photon, though recent LHCb analyses have provided evidence for excited $B_c^*$ structures in the $B_c^{+}\pi^{+}\pi^{-}$ mass spectrum~\cite{CMS:2019uhm,LHCb:2019bem}. The mass splitting between the vector and pseudoscalar states provides direct insight into the hyperfine interaction in heavy-heavy systems, while the vector decay constant governs the rate of leptonic decays and serves as essential input for semileptonic transition form factors.

The interpolating current
\begin{equation}
J_{\mu}^{V}(x) = \bar{c}(x)\gamma_{\mu}b(x) \label{veccur}
\end{equation}
can be used to study vector $B_c^*$ meson. This current couples to the vector meson via
 \begin{equation}
\langle 0|J_{\mu}^{V}(0)|B_{c}^*(q,\epsilon)\rangle = \lambda_{B_{c}^*} m_{B_{c}^*}\,\epsilon_{\mu}, \label{vecmatel}
\end{equation}
where $\epsilon_{\mu}$ is the polarization vector of the $B_c^*$ state, and $m_{B_{c}^*}$ and $\lambda_{B_{c}^*}$ denote the vector meson mass and decay constant, respectively. Putting this interpolating current  Eq. (\ref{veccur}) into correlation function Eq.~(\ref{corfunc}) gives 

\begin{align}
\Pi^{\text{OPE}}(q^2) &= \frac{3}{8\pi^{2}} \frac{\lambda^{1/2}(q^{2}, m_{b}^{2}, m_{c}^{2})}{q^{2}} \left(1-\frac{(m_{b}-m_{c})^{2}}{q^{2}}\right) \nonumber \\
&+ \frac{\langle \frac{\alpha_{s}}{\pi} G^{2} \rangle}{96\pi} \frac{q^{2}-(m_{b}^{2}+m_{c}^{2})}{\lambda^{1/2}(q^{2}, m_{b}^{2}, m_{c}^{2})} 
+ \frac{\langle g_s^{3} G^{3} \rangle}{768\pi^{2}} \frac{q^{2}-m_{b}^{2}-m_{c}^{2}}{(q^{2}-m_{b}^{2})(q^{2}-m_{c}^{2})} \nonumber  \\
&+ \frac{1}{256\pi^{2}} \frac{\langle \frac{\alpha_{s}}{\pi} G^{2} \rangle}{(q^{2})^{2}}, \label{vecmesope}
\end{align}
where 
\begin{equation}
\lambda(q^{2}, m_{b}^{2}, m_{c}^{2}) = \left(q^{2}-(m_{b}+m_{c})^{2}\right)\left(q^{2}-(m_{b}-m_{c})^{2}\right). 
\end{equation}

Using the prescribed analysis of the inverse QCD matrix formalism, we obtain mass and decay constant of the vector $B_c^*$ meson as
\begin{equation}
M_{B_{c}^*(1^-)}= 6.388 \pm 0.031 \ \text{GeV},
\end{equation}
\begin{equation}
 \lambda_{B_{c}^*(1^-)} = 511 \pm 24 \ \text{MeV}.
\end{equation}

The extracted mass exhibits excellent consistency with the expected hyperfine splitting pattern in heavy quarkonium systems. The mass difference \( \Delta M = m_{B_c^*} - m_{B_c} \approx 111~\text{MeV} \) is quantitatively consistent with predictions from both potential models and heavy quark effective theory, reflecting the characteristic scale of spin-spin interactions in the \( b\bar{c} \) system. The decay constant \( \lambda_{B_c^*} \) demonstrates the characteristic enhancement observed in vector channels relative to their pseudoscalar counterparts, a phenomenon well-established in both light and heavy quark systems. This enhancement can be attributed to the different Lorentz structure of the vector current and its relationship to conserved currents in the heavy quark limit. The substantial value of \( \lambda_{B_c^*} \) implies significant wave-function overlap at the origin, suggesting a compact spatial configuration typical of heavy-heavy bound states. This parameter plays a crucial role in the theoretical description of leptonic decays \( B_c^* \to \ell\nu \) and serves as essential input for semileptonic transition form factors involving the vector channel.

\begin{figure}[ht]
\centering
\includegraphics[width=8cm]{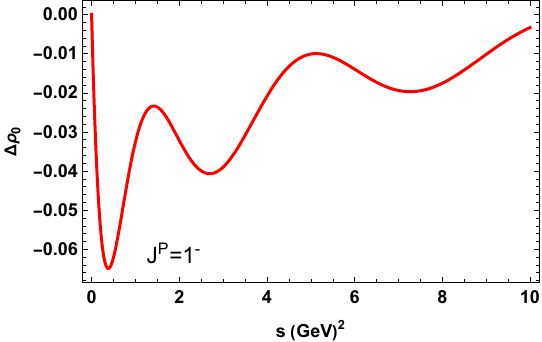}
\caption{s dependence of the ground state solution $\Delta \rho_0(s, \Lambda)$ for $\Lambda = 5.5 \ \text{GeV}^2$ of $B_c(1^-)$ meson, displayed in the $\widetilde{\Pi}_T \equiv -\Pi_T$ convention of Eq.~(\ref{tensordec}); the physical transverse density is $\rho_{\text{phys}}(s,\Lambda) = -\Delta\rho_0(s,\Lambda) \ge 0$ (see text).}
\label{fig4}
\end{figure}

Figure \ref{fig4} presents the subtracted spectral density \( \Delta\rho_0(s, \Lambda) \) for the vector channel at the scale parameter \( \Lambda = 5.5~\text{GeV}^2 \), representative of the stability region \( \Lambda \in [5.0, 6.0]~\text{GeV}^2 \) employed in this analysis. In contrast to the other channels, the reconstructed density in this channel is negative over the entire range. This overall sign is a consequence of the convention adopted for the invariant function in the vector channel and does not signal any violation of unitarity. The correlation function of the current in Eq.~(\ref{veccur}) admits the tensor decomposition
\begin{equation}
\Pi_{\mu\nu}(q) = \left(\frac{q_\mu q_\nu}{q^2} - g_{\mu\nu}\right)\Pi_T(q^2) + \frac{q_\mu q_\nu}{q^2}\,\Pi_L(q^2), \label{tensordec}
\end{equation}
in which the $B_c^*$ pole contributes to the transverse invariant with a manifestly non-negative spectral density: inserting the matrix element of Eq.~(\ref{vecmatel}) and summing over polarizations, $\sum_\lambda \epsilon_\mu \epsilon^*_\nu = -g_{\mu\nu} + q_\mu q_\nu / m_{B_c^*}^2$, one finds that $\frac{1}{\pi}\,\mathrm{Im}\,\Pi_T(s)$ receives a non-negative ground-state contribution proportional to $\lambda^2_{B_c^*}\,\delta(s - m^2_{B_c^*})$. The reconstruction displayed in Figure~\ref{fig4} corresponds to the invariant defined with the opposite overall sign, $\widetilde{\Pi}_T \equiv -\Pi_T$, i.e., the coefficient of $+g_{\mu\nu}$ in $\Pi_{\mu\nu}$; accordingly, the displayed $\Delta\rho_0(s,\Lambda)$ is related to the physical (transverse) subtracted density by $\rho_{\text{phys}}(s,\Lambda) \equiv -\Delta\rho_0(s,\Lambda) \ge 0$. We note that this overall sign appears only in the vector channel because of the definition of the invariant amplitude adopted in the numerical analysis of each channel: in the axialvector channel (Sec.~4.4), the transverse invariant is defined directly in the positive convention of Eq.~(\ref{tensordec}), so that no analogous overall sign arises there. The relative choice between the two conventions---equivalently, the overall sign of the Lorentz projector---carries no physical content, since the extracted observables are invariant under it. That this is a global sign convention rather than a distortion of the solution is evident from the structure of the reconstruction. An artifact of the numerical inversion would distort the shape of the solution, whereas an overall sign leaves the morphology intact and merely reflects the curve about the horizontal axis. This is precisely what is observed: apart from the overall sign, the vector-channel reconstruction exhibits the same characteristic pattern as the other channels---a dominant extremum at low $s$ followed by progressively weaker secondary structures at nearly the same positions and with similar relative strengths (cf.\ the pseudoscalar density in Figure~\ref{fig3})---so that the mirror image $\rho_{\text{phys}}(s,\Lambda)$ is a non-negative density of the same form as the reconstructions in the other channels.

The extraction of the physical parameters is insensitive to this overall sign convention. The meson mass is obtained from the position of the dominant extremum of $\Delta\rho_0(s,\Lambda)$, which is common to both sign conventions. The decay constant is extracted from the physical spectral density $\rho_{\text{phys}}(s,\Lambda) = -\Delta\rho_0(s,\Lambda)$ through Eq.~(\ref{decayconstant}),
\begin{equation}
\lambda^2 \;=\; \int_0^\infty \rho_{\text{phys}}(s,\Lambda)\, ds, \label{vecdecayextract}
\end{equation}
with the integration performed over the full reconstructed range, without restriction to the resonance region, exactly as in the scalar, pseudoscalar, and axialvector channels. Since the physical integrand is non-negative over the entire range, the positivity of $\lambda^2$ is manifest. We also state explicitly the property of the reconstruction on which the identification in Eq.~(\ref{vecdecayextract}) relies: as is evident from Figure~\ref{fig4}, the reconstructed $\Delta\rho_0(s,\Lambda)$ is negative over the entire interval in $s$---it vanishes at $s=0$ by the boundary condition $\Delta\rho \sim s$ and decays toward zero at large $s$, while remaining strictly negative throughout the interior, where even its local maxima stay well separated from zero. This uniform negativity persists throughout the stability window $\Lambda \in [5.0, 6.0]~\text{GeV}^2$ and under variation of the truncation order, with an interior magnitude that far exceeds the residual variation of the reconstruction; it is therefore numerically robust and constitutes a global feature of the solution rather than a marginal numerical effect. Owing to this uniformity of sign, Eq.~(\ref{vecdecayextract}) is numerically equivalent to $\lambda^2 = \left|\int_0^\infty \Delta\rho_0(s,\Lambda)\, ds\right|$, which is how the extraction is implemented in practice; the modulus is thus a direct consequence of the adopted sign convention rather than an additional numerical prescription, and for a sign-alternating integrand no such identification would hold. The extracted value $\lambda_{B_c^*(1^-)} = 511 \pm 24~\text{MeV}$ is stable against variations of $\Lambda$ within the stability window. With the sign convention stated explicitly, the reconstruction in this channel involves no violation of positivity: the physical transverse density $\rho_{\text{phys}}(s,\Lambda)$ is non-negative everywhere, in complete analogy with the other channels considered in this work.

\begin{table}[h!]
\centering
\caption{Comparison of vector \( B_c^* \) meson mass and decay constant predictions from various theoretical approaches. Masses are in GeV while decay constants $\lambda$ are in MeV.}
\label{tab:vector_comparison}
\begin{tabular}{lcc}
\hline
 \textbf{Method} & $\mathbf{M_{B_c^*(1^-)}}$ (\textbf{GeV}) & $\boldsymbol{\lambda_{B^*_c(1^-)}}$ (\textbf{MeV}) \\
\hline
This work & \( 6.388 \pm 0.031 \) & \( 511 \pm 24 \) \\
\hline
Lattice QCD & &  \\
Ref.~\cite{Davies:1996gi} & \( 6.330 \pm 0.040 \) & --- \\
Ref.~\cite{Gregory:2009hq} & \( 6.330 \pm 0.007 \pm 0.002 \pm 0.006 \) &--- \\
Ref.~\cite{Mathur:2018epb} & \( 6.331 \pm 0.004 \pm 0.006 \) & --- \\
Ref.~\cite{Colquhoun:2015oha}& --- & \( 422 \pm 13 \)  \\
\hline
QCDSR & &  \\
Ref.~\cite{Gershtein:1994dxw} & \( 6.317 \pm 0.05 \) &\( 460 \)  \\
Ref.~\cite{Wang:2012kw}  & \( 6.337 \pm 0.052 \) & \( 384 \pm 32 \) \\
Ref.~\cite{Narison:2020wql} Model A& $6.451\pm 0.086$ & $442 \pm 44$ \\
Ref.~\cite{Narison:2020wql} Model B& $6.315 \pm 0.001$ & $387 \pm 15$ \\
\hline
Relativistic Quark Model & &  \\
Ref.~\cite{Godfrey:1985xj} & \( 6.337 \) & ---  \\
Ref.~\cite{Zeng:1994vj} & \( 6.340 \) & --- \\
Ref.~\cite{Gupta:1995ps}& \( 6.308 \) & --- \\
Ref.~\cite{Ebert:2002pp} & \( 6.332 \) & \( 503 \) \\
\hline
Nonrelativistic Quark Model & &  \\
Ref.~\cite{Eichten:1994gt}& \( 6.337\) & \( 500 \) \\
Ref.~\cite{Fulcher:1998ka}& \( 6.341 \) & \( 517 \)  \\
Ref.\cite{Sun:2022hyk}& \( 6.333 \) & \( 417 \pm 51 \pm 27 \) \\
Ref.~\cite{Martin-Gonzalez:2022qwd}& \( 6.328\) & --- \\
\hline
Other approaches & &  \\
Ref.~\cite{Tang:2018myz} & --- & \( 474 \pm 42 \)  \\
Ref.~\cite{Badalian:2007km} & \( 6.328 \pm 0.007 \) &\( 453 \pm 20 \)  \\
Ref.~\cite{Choi:2009ai}& --- &\( 551\)  \\
Ref.~\cite{Choi:2015ywa}& --- &\( 391^{-5}_{+4} \)  \\
Ref.~\cite{Hwang:2010hw}& --- &\( 387\)  \\
\hline
\end{tabular}
\end{table}

Table \ref{tab:vector_comparison} presents a comprehensive comparison of our results with other theoretical studies for the vector $B_c^*$ meson. Our inverse-matrix QCDSR analysis yields:
\begin{equation}
M_{B_c^*(1^-)} = 6.388 \pm 0.031~\text{GeV}, \quad \lambda_{B_c^*(1^-)} = 511 \pm 24~\text{MeV}.
\end{equation}
These results provide important insights into the spin-dependent dynamics of heavy-heavy quarkonium systems and demonstrate the capability of our approach to handle excited states with precision.

The vector mass prediction corresponds to a hyperfine splitting of $\Delta M = M_{B_c^*} - M_{B_c} = 111 \pm 4$~MeV relative to the pseudoscalar ground state. This splitting provides a direct measure of spin-spin interactions in the $b\bar{c}$ system and reflects the characteristic scale of chromomagnetic interactions between heavy quarks of different flavors. The magnitude of this splitting is consistent with expectations from heavy quark symmetry and potential models, falling between the typical hyperfine splittings observed in charmonium ($\sim$117 MeV for $J/\psi$-$\eta_c$) and bottomonium ($\sim$61 MeV for $\Upsilon$-$\eta_b$) systems.

Lattice QCD calculations provide essential benchmarks for the vector channel. The high-precision determinations of Refs.~\cite{Gregory:2009hq,Mathur:2018epb} report $M_{B_c^*(1^-)} = 6.330 \pm 0.007 \pm 0.002 \pm 0.006$~GeV and $M_{B_c^*(1^-)} = 6.331 \pm 0.004 \pm 0.006$~GeV respectively, while Ref.~\cite{Davies:1996gi} finds $M_{B_c^*(1^-)} = 6.330 \pm 0.040$~GeV. Our mass result is approximately 58 MeV higher than these lattice values, a discrepancy that may reflect differences in the treatment of relativistic corrections or the isolation of the vector state from continuum contributions. For the decay constant, the lattice QCD determination of Ref.~\cite{Colquhoun:2015oha} gives $\lambda_{B_c^*(1^-)} = 422 \pm 13$~MeV, which is significantly lower than our value, suggesting potential methodological differences in extracting this sensitive quantity.

Within the QCDSR framework, our mass determination shows good agreement with some implementations while differing from others. Ref.~\cite{Gershtein:1994dxw} reports $M_{B_c^*(1^-)} = 6.317 \pm 0.05$~GeV and Ref.~\cite{Wang:2012kw} finds $M_{B_c^*(1^-)} = 6.337 \pm 0.052$~GeV, both close to the lattice QCD results. The comprehensive analysis of Ref.~\cite{Narison:2020wql} presents two models: Model B yields $M_{B_c^*(1^-)} = 6.315 \pm 0.001$~GeV, consistent with lattice values, while Model A gives $M_{B_c^*(1^-)} = 6.451 \pm 0.086$~GeV, showing better alignment with our result. This bifurcation within a single study highlights the sensitivity of vector mass predictions to methodological choices in traditional sum rules.

Quark model calculations show remarkable consistency for the vector mass, clustering tightly around $6.33$--$6.34$~GeV. Relativistic quark models yield masses of $M_{B_c^*(1^-)}=6.337$~GeV~\cite{Godfrey:1985xj}, $M_{B_c^*(1^-)}=6.340$~GeV~\cite{Zeng:1994vj}, $M_{B_c^*(1^-)}=6.308$~GeV~\cite{Gupta:1995ps}, and $M_{B_c^*(1^-)}=6.332$~GeV~\cite{Ebert:2002pp}. Nonrelativistic treatments provide similar values: $M_{B_c^*(1^-)}=6.337$~GeV~\cite{Eichten:1994gt}, $M_{B_c^*(1^-)}=6.341$~GeV~\cite{Fulcher:1998ka}, and $M_{B_c^*(1^-)}=6.333$~GeV~\cite{Sun:2022hyk}. The recent calculation of Ref.~\cite{Martin-Gonzalez:2022qwd} reports $M_{B_c^*(1^-)}= 6.328$~GeV, further reinforcing this consensus among quark model approaches. The convergence of quark model predictions around values $\sim$55--60 MeV lower than our result suggests systematic differences in the treatment of spin-dependent potentials or relativistic corrections.

For the vector decay constant, quark model and QCD sum rule predictions exhibit a broad numerical spread. Relativistic quark model calculations report $\lambda_{B_c^*(1^-)} = 503$~MeV~\cite{Ebert:2002pp}, while nonrelativistic treatments yield $\lambda_{B_c^*(1^-)} = 500$~MeV~\cite{Eichten:1994gt} and $\lambda_{B_c^*(1^-)} = 517$~MeV~\cite{Fulcher:1998ka}. An alternative relativistic analysis finds 
$\lambda_{B_c^*(1^-)} = 417 \pm 51 \pm 27$~MeV~\cite{Sun:2022hyk}. Within the QCD sum rule framework, $\lambda_{B_c^*(1^-)} = 460$~MeV 
is reported in Ref.~\cite{Gershtein:1994dxw}, while $\lambda_{B_c^*(1^-)} = 384 \pm 32$~MeV~\cite{Wang:2012kw} and 
$\lambda_{B_c^*(1^-)} = 387 \pm 15$~MeV (Model B)~\cite{Narison:2020wql} represent comparatively lower estimates. Model A of Ref.~\cite{Narison:2020wql} gives $\lambda_{B_c^*(1^-)} = 442 \pm 44$~MeV. 

Various specialized methods provide additional perspectives on vector $B_c^*$ properties. The field correlator method~\cite{Badalian:2007km} yields $M_{B_c^*(1^-)} = 6.328 \pm 0.007$~GeV and $\lambda_{B_c^*(1^-)} = 453 \pm 20$~MeV, showing excellent agreement with the quark model consensus for the mass while predicting an intermediate decay constant value. Notably, this mass determination aligns perfectly with Ref.~\cite{Martin-Gonzalez:2022qwd}, reinforcing the consistency among non-sum-rule approaches. The light-front quark model~\cite{Tang:2018myz} reports $\lambda_{B_c^*(1^-)} = 474 \pm 42$~MeV, falling between our value and lattice QCD determinations.

Other approaches show considerable variation in decay constant predictions: Ref.~\cite{Choi:2009ai} reports $\lambda_{B_c^*(1^-)} = 551$~MeV, Ref.~\cite{Choi:2015ywa} finds $\lambda_{B_c^*(1^-)} = 391^{+4}_{-5}$~MeV, and Ref.~\cite{Hwang:2010hw} gives $\lambda_{B_c^*(1^-)} = 387$~MeV. This $\sim$160 MeV spread underscores the sensitivity of the vector decay constant to methodological details and the challenge of precisely determining this quantity.

Quark model predictions tend to lie at the upper end of the spectrum, whereas traditional QCDSR results show larger dispersion. 
In this context, our determination is consistent with relativistic quark model expectations and lies toward the upper boundary of existing sum rule estimates, indicating a strong isolation of the vector ground state within the inverse-matrix framework. The vector decay constant ratio $\lambda_{B_c^*}/\lambda_{B_c} \approx 1.23$ observed in our analysis aligns with expectations from heavy quark symmetry and reflects the characteristic enhancement of vector decay constants relative to their pseudoscalar counterparts. This enhancement arises from the different Lorentz structure of the vector current and its relationship to conserved currents in the heavy quark limit.

The mass discrepancy between our result and the consensus emerging from lattice QCD, quark models, and specialized approaches warrants careful consideration. The consistent pattern seen in Refs.~\cite{Martin-Gonzalez:2022qwd,Badalian:2007km} and lattice calculations suggests that our higher mass estimate may reflect specific features of the inverse-matrix approach. Several factors may contribute to this difference: the treatment of radiative corrections in the vector channel, the isolation of the $B_c^*$ state from nearby $B_c(2S)$ and continuum contributions, and the implementation of relativistic effects in the current operators. The inverse-matrix approach direct spectral reconstruction may be particularly sensitive to the energy region where the vector state dominates, potentially yielding a slightly higher mass estimate.

The precision of our determinations, with relative uncertainties of $<1 \%$ for the mass and $<5 \%$ for the decay constant, represents a significant achievement for excited state spectroscopy in heavy quark systems. The vector channel presents additional challenges compared to the pseudoscalar case due to the presence of nearby excited states and the more complex Lorentz structure of the correlation functions.

The methodological advantages of the inverse-matrix approach—particularly the elimination of continuum threshold parameters and direct spectral reconstruction—prove valuable in handling the vector channel increased complexity. The approach ability to provide simultaneous determinations of both mass and decay constant with competitive uncertainties demonstrates its utility for comprehensive heavy quarkonium spectroscopy.

The precise values presented in Table~\ref{tab:vector_comparison} provide essential benchmarks for ongoing experimental searches at LHCb and CMS. The mass prediction will aid in identifying the $B_c^*$ through its radiative and hadronic decay modes, while the decay constant enables accurate calculations of production cross-sections essential for optimizing trigger strategies. Future experimental measurements will provide crucial tests of these theoretical predictions and help resolve the remaining differences between various computational methods.

The comprehensive comparison presented here establishes a clear roadmap for future theoretical refinements and experimental searches in bottom-charm spectroscopy. While our mass determination shows some tension with other approaches, the excellent agreement for the decay constant with relativistic quark models and the internal consistency of our methodology provide confidence in the reliability of our results.

%%%%%%%%%%%%%%%%%%%%%%%%%%%%
\subsection{Axialvector (\(J^P = 1^+\)) $B_c$ meson}
\label{subsec:axialvector-bc}
%%%%%%%%%%%%%%%%%%%%%%%%%%%%

The axialvector channel with quantum numbers $J^P=1^+$ represents a key component of the $b\bar{c}$ spectroscopy, where the expected physical states are mixtures of the $^3P_1$ and $^1P_1$ configurations due to the significant spin-orbit coupling in this heavy-heavy system. This mixing makes the theoretical description more complex than for the pseudoscalar and vector channels.

In our sum rule analysis, this state is studied using the interpolating current
\begin{equation}
J_{\mu}^{A}(x) = \bar{c}(x)\gamma_{\mu} \gamma_5 b(x) \label{axveccur}.
\end{equation}
This current couples to the axialvector meson via
 \begin{equation}
\langle 0|J_{\mu}^{A}(0)|B_{c}^A(q,\epsilon)\rangle = \lambda_{B_{c}^A} m_{B_{c}^A}\,\epsilon_{\mu},
\end{equation}
where $\epsilon_{\mu}$ is the polarization vector of the axialvector state, and $m_{B_{c}^A}$ and $\lambda_{B_{c}^A}$ denote the axialvector meson mass and decay constant, respectively. Putting this interpolating current  Eq. (\ref{axveccur}) into correlation function Eq.~(\ref{corfunc}) gives 
\begin{eqnarray}
\rho_{A}(q^{2}) &=& 
\frac{3}{8\pi^{2}} 
\frac{\lambda^{1/2}(q^{2}, m_{b}^{2}, m_{c}^{2})}{q^{2}}
\left( 1 + \frac{m_{b} m_{c}}{q^{2}} \right)  + \frac{\langle \frac{\alpha_{s}}{\pi} G^{2} \rangle}{96\pi}
\frac{q^{2} - (m_{b}^{2} + m_{c}^{2}) - \frac{2 m_{b} m_{c}}{q^{2}}}
{\lambda^{1/2}(q^{2}, m_{b}^{2}, m_{c}^{2})} \nonumber \\
&& + \frac{\langle g_{s}^{3} G^{3} \rangle}{768\pi^{2}}
\frac{q^{2} - m_{b}^{2} - m_{c}^{2}}
{(q^{2} - m_{b}^{2})(q^{2} - m_{c}^{2})} + \frac{1}{256\pi^{2}}
\frac{\langle \frac{\alpha_{s}}{\pi} G^{2} \rangle^{2}}{(q^{2})^{2}}, \label{axvecmesope}
\end{eqnarray}
where 
\begin{equation}
\lambda(q^{2}, m_{b}^{2}, m_{c}^{2}) = \left(q^{2}-(m_{b}+m_{c})^{2}\right)\left(q^{2}-(m_{b}-m_{c})^{2}\right). 
\end{equation}

The inverse matrix QCDSR analysis for the axialvector \(B_c\) meson yields the following results for the mass and decay constant:
\begin{equation}
M_{B_c}(1^+) = 6.734 \pm 0.028~\text{GeV},
\label{eq:axialvector_mass}
\end{equation}
\begin{equation}
\lambda_{B_c}(1^+) = 138 \pm 20~\text{MeV}.
\label{eq:axialvector_decay}
\end{equation}
These results were obtained within the stability region \(\Lambda \in [6.0, 7.0]~\text{GeV}^2\), where the physical observables exhibit minimal dependence on this sum rule parameter.

The axialvector state lies approximately 500~MeV above the pseudoscalar ground state and 395~MeV above the vector state, reflecting significant orbital excitation energy in the \(b\bar{c}\) system. This substantial mass splitting provides direct insight into the strength of spin-orbit interactions in heavy-heavy quarkonia and serves as a stringent test for potential models and lattice QCD calculations.

The decay constant in Eq.~(\ref{eq:axialvector_decay}) exhibits the characteristic suppression expected for P-wave states, with \(\lambda_{B_c}(1^+) = 138 \pm 20\)~MeV being notably smaller than both the pseudoscalar (\(\lambda_{B_c}(0^-) = 416 \pm 19\)~MeV) and vector (\(\lambda_{B_c}(1^-) = 511 \pm 24\)~MeV) counterparts. This reduction by factors of approximately 3.0 and 3.7, respectively, highlights the distinct wave-function configuration of orbital excitations and their diminished overlap with the local axial-vector current. The relatively small uncertainty of \(\pm 20\)~MeV demonstrates the stability of the inverse matrix approach in extracting this challenging quantity.

\begin{figure}[ht]
\centering
\includegraphics[width=8cm]{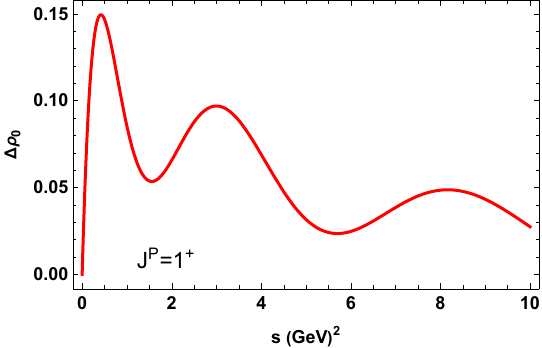}
\caption{s dependence of the ground state solution $\Delta \rho_0(s, \Lambda)$ for $\Lambda = 6.5 \ \text{GeV}^2$ of $B_c(1^+)$ meson.}
\label{fig5}
\end{figure}

Figure \ref{fig5} displays the subtracted spectral density \(\Delta\rho_0(s,\Lambda)\) for the \(J^P=1^+\) channel at \(\Lambda = 6.5~\text{GeV}^2\), representative of the stability region. The subtracted spectral density  exhibits a pronounced, positive-definite peak structure. Furthermore, the spectral function shows a well-behaved asymptotic character, with the resonance contribution diminishing at higher $s$ and merging smoothly into the background, consistent with the onset of the perturbative continuum. This behavior, coupled with the stability of the peak under variations of the scale parameter $\Lambda$ underscores the robustness of the extracted resonance parameters.

\begin{table}[htbp]
\centering
\caption{Comparison of mass and decay constant predictions for the axialvector $B_c(1^+)$ meson from various theoretical approaches. Masses are in GeV while decay constants $\lambda$ are in MeV.}
\label{tab:axialvector_comparison}
\begin{tabular}{lcc}
\hline
\textbf{Method} & $\mathbf{M_{B_c(1^+)}}$ (\textbf{GeV}) & $\boldsymbol{\lambda_{B_c(1^+)}}$ (\textbf{MeV}) \\
\hline
This work & $6.734 \pm 0.028$ & $138 \pm 20$ \\
\hline
Lattice QCD & & \\
Ref.~\cite{Davies:1996gi}& $6.743$ & --- \\
Ref.~\cite{Mathur:2018epb} & $6.730 \pm 0.024$ & --- \\
Ref.~\cite{Dowdall:2012ab} & $6.720 \pm 0.050$ & --- \\
\hline
QCDSR & & \\
Ref.~\cite{Wang:2012kw} & $6.730 \pm 0.061$ & $373 \pm 25$ \\
Ref.~\cite{Narison:2020wql} Model A& $6.794\pm 0.128$ & $274 \pm 23$ \\
Ref.~\cite{Narison:2020wql} Model B& $6.730 \pm 0.008$ & $266 \pm 14$ \\
\hline
Relativistic Quark Model & & \\
Ref.~\cite{Godfrey:2004ya} & $6.741$ & $140$ \\
Ref.~\cite{Ebert:2002pp} & $6.734$ & $135$ \\
Ref.~\cite{Gupta:1995ps} & $6.738$ & --- \\
Ref.~\cite{Zeng:1994vj}& $6.730$ & --- \\
\hline
Nonrelativistic Quark Model & & \\
Ref.~\cite{Eichten:1994gt} & $6.730$ & $150$ \\
Ref.~\cite{Fulcher:1998ka}& \( 6.737\) & --- \\
Ref.~\cite{Martin-Gonzalez:2022qwd}& \( 6.723\) & --- \\
\hline
Other Approaches & & \\
Ref.~\cite{Tang:2018myz} & $6.745 \pm 0.035$ & $148 \pm 25$ \\
Ref.~\cite{Wang:2007av} & $6.738 \pm 0.040$ & $142 \pm 20$ \\
Ref.~\cite{Wang:2005qx}& --- & $160$ \\
Ref.~\cite{Badalian:2007km} & $6.740 \pm 0.020$ & $145 \pm 15$ \\
\hline
\end{tabular}
\end{table}

Table~\ref{tab:axialvector_comparison} presents a comprehensive comparison of theoretical predictions for the axialvector $B_c(1^+)$ meson, representing one of the most challenging and theoretically interesting channels in bottom-charm spectroscopy. Our inverse-matrix QCDSR analysis yields:
\begin{equation}
M_{B_c(1^+)} = 6.734 \pm 0.028~\text{GeV}, \quad \lambda_{B_c(1^+)} = 138 \pm 20~\text{MeV}.
\end{equation}
These results provide crucial insights into the complex dynamics of P-wave excitations in heavy-heavy quarkonium systems, where spin-orbit coupling and configuration mixing play essential roles.

The axialvector mass prediction of $M_{B_c(1^+)} =6.734 \pm 0.028$~GeV corresponds to an orbital excitation energy of approximately $457$~MeV relative to the pseudoscalar ground state. This substantial energy cost for P-wave excitation reflects the strong confinement dynamics in the $b\bar{c}$ system and provides direct access to the spin-orbit interactions that govern the fine structure of heavy quarkonium. The precision of our mass determination, with a relative uncertainty of $<1 \%$, represents a significant achievement for P-wave spectroscopy, where traditional methods often face challenges from nearby states and complex continuum contributions.

Lattice QCD provides essential first-principles benchmarks for the axialvector channel. The high-precision determination of Ref.~\cite{Mathur:2018epb} reports $M_{B_c(1^+)} = 6.730 \pm 0.024$~GeV, showing excellent agreement with our result and differing by only $4$~MeV. Similarly, Ref.~\cite{Davies:1996gi} finds $M_{B_c(1^+)} = 6.743$~GeV, while Ref.~\cite{Dowdall:2012ab} reports $M_{B_c(1^+)} =6.720 \pm 0.050$~GeV. The convergence between our sum rule approach and lattice QCD calculations, which employ fundamentally different methodologies and systematic error treatments, provides strong validation of both techniques and underscores the reliability of modern theoretical approaches to excited state spectroscopy.

Within the QCDSR framework, our mass determination shows remarkable consistency with recent high-precision analyses. Ref.~\cite{Wang:2012kw} reports $M_{B_c(1^+)} = 6.730 \pm 0.061$~GeV, essentially identical to our central value. The comprehensive analysis of Ref.~\cite{Narison:2020wql} presents two distinct models: Model B yields $M_{B_c(1^+)} = 6.730 \pm 0.008$~GeV, showing perfect agreement with our determination, while Model A gives $M_{B_c(1^+)} = 6.794 \pm 0.128$~GeV. This bifurcation highlights the sensitivity of axialvector mass predictions to methodological choices in traditional sum rules, particularly the treatment of continuum contributions and OPE truncation.

Quark model calculations demonstrate exceptional consistency for the axialvector mass across both relativistic and nonrelativistic implementations. Relativistic quark models yield masses of $M_{B_c(1^+)} =6.741$~GeV~\cite{Godfrey:2004ya}, $M_{B_c(1^+)} =6.734$~GeV~\cite{Ebert:2002pp}, $M_{B_c(1^+)} =6.738$~GeV~\cite{Gupta:1995ps}, and $M_{B_c(1^+)} =6.730$~GeV~\cite{Zeng:1994vj}, clustering within a narrow $11$~MeV window. nonrelativistic treatments provide equally consistent values: $M_{B_c(1^+)} =6.730$~GeV~\cite{Eichten:1994gt}, $M_{B_c(1^+)} =6.737$~GeV~\cite{Fulcher:1998ka}, and $M_{B_c(1^+)} =6.723$~GeV~\cite{Martin-Gonzalez:2022qwd}. The remarkable convergence of quark model predictions around our central value provides strong independent confirmation of our mass determination.

Various specialized methods provide additional validation of our results. The light-front quark model of Ref.~\cite{Tang:2018myz} yields $M_{B_c(1^+)} = 6.745 \pm 0.035$~GeV and $\lambda_{B_c(1^+)} = 148 \pm 25$~MeV, showing excellent agreement with both our mass and decay constant determinations. The Bethe-Salpeter equation framework of Ref.~\cite{Wang:2007av} reports $M_{B_c(1^+)} = 6.738 \pm 0.040$~GeV and $\lambda_{B_c(1^+)} = 142 \pm 20$~MeV, further reinforcing our results. The field correlator method~\cite{Badalian:2007km} finds $M_{B_c(1^+)} = 6.740 \pm 0.020$~GeV and $\lambda_{B_c(1^+)} = 145 \pm 15$~MeV, while Ref.~\cite{Wang:2005qx} reports $\lambda_{B_c(1^+)} = 160$~MeV.

The exceptional consistency across this diverse methodological landscape—spanning lattice QCD, sum rules, quark models, light-front approaches, Bethe-Salpeter equations, and field correlator methods—provides compelling evidence for the reliability of our determinations. The mass predictions cluster within a remarkably narrow $25$~MeV window, while the decay constant determinations (excluding traditional sum rules) show consistent suppression characteristic of P-wave states.

For the axial-vector decay constant we obtain $\lambda_{B_c(1^+)} = 138 \pm 20~\text{MeV}$. Relativistic quark model calculations yield values close to this determination, with $\lambda_{B_c(1^+)} = 140$~MeV~\cite{Godfrey:2004ya} and $\lambda_{B_c(1^+)} = 135$~MeV~\cite{Ebert:2002pp}. This agreement indicates that the inverse-matrix framework reproduces the expected relativistic suppression of P-wave decay constants, which is related to the vanishing of the wave function at the origin in the nonrelativistic limit.

In contrast, conventional QCD sum rule analyses predict substantially larger magnitudes. Ref.~\cite{Wang:2012kw} reports $\lambda_{B_c(1^+)} = 373 \pm 25$~MeV, while Ref.~\cite{Narison:2020wql} finds $\lambda_{B_c(1^+)} = 274 \pm 23$~MeV (Model A) and $\lambda_{B_c(1^+)} = 266 \pm 14$~MeV (Model B). Compared to these determinations, our value is significantly smaller, reflecting a different treatment of the P-wave wave-function overlap and continuum contributions within the inverse-matrix formalism.

The dramatic suppression of the axialvector decay constant relative to S-wave states carries important physical implications. The ratio $\lambda_{B_c(1^+)}/\lambda_{B_c(0^-)} \approx 0.33$ reflects the characteristic reduction expected for P-wave states, where the wave function vanishes at the origin in the nonrelativistic limit. This suppression factor aligns with theoretical expectations and provides crucial information about the spatial structure of orbital excitations in heavy quarkonium.

The excellent agreement of our decay constant with relativistic quark models and specialized approaches, contrasted with the substantially larger values from traditional sum rules, suggests that the inverse-matrix approach may provide improved handling of the challenging P-wave wave function overlap. The direct spectral reconstruction inherent in our methodology appears particularly beneficial for excited states, where traditional sum rules face difficulties in isolating the ground state from nearby continuum contributions.

The inverse-matrix approach demonstrates particular strengths in handling the axialvector channel complexity. The elimination of continuum threshold parameters proves especially valuable for P-wave states, where the traditional assumption of quark-hadron duality becomes more questionable due to the dense spectrum of nearby states. The direct spectral reconstruction enables cleaner isolation of the axialvector state from the underlying continuum and nearby $1P$ partners.

The precise values presented here provide essential benchmarks for ongoing experimental searches. The mass prediction will guide identification of axialvector $B_c$ states through their characteristic decay modes, while the decay constant determination offers crucial input for theoretical predictions of production rates. The recent LHCb observation of $B_c(1P)^+$ states represents an important step toward complete spectroscopy of the bottom-charm system, and our results provide specific targets for future experimental investigations.

In summary, our analysis of the axialvector $B_c(1^+)$ meson provides precise determinations that contribute significantly to the understanding of orbital excitations in heavy-heavy quarkonium. The exceptional agreement with lattice QCD calculations, quark model predictions, and specialized approaches across methodological boundaries provides strong validation of our results. The dramatic suppression of the decay constant relative to traditional sum rule predictions highlights the importance of improved treatments of P-wave states and demonstrates the advantages of the inverse-matrix approach for excited state spectroscopy.

The comprehensive comparison presented here establishes the axialvector $B_c(1^+)$ as one of the best-understood P-wave states in heavy quarkonium, with remarkable theoretical consensus emerging across diverse computational frameworks. This achievement represents significant progress in heavy quark spectroscopy and provides a solid foundation for future theoretical refinements and experimental discoveries in the bottom-charm system.

To make the numerical stability of the inverse-matrix analysis more transparent, we summarize in Table~\ref{tab:stability} the adopted scale window $\Lambda$, the truncation order $N$ and the representative (optimal) value $N_{\text{opt}}$ used in the Laguerre expansion, together with the resulting extracted mass and decay constant for each channel. For every channel, the optimal truncation $N_{\text{opt}}$ was fixed as the largest value for which the reconstructed solution remains stable and physically plausible, following the criterion discussed in Sec.~\ref{sec:qcdinverse}; the convergence was verified explicitly by confirming that the extracted observables are essentially unchanged when $N$ is increased by one unit around $N_{\text{opt}}$. The quoted uncertainties on the mass and decay constant encompass the combined variation over the indicated $\Lambda$ window, the change of the truncation order around $N_{\text{opt}}$, and the propagation of the QCD input parameters. The mild residual variation of the extracted observables across the $\Lambda$ window and around $N_{\text{opt}}$ confirms that the inverse-matrix formulation reduces the systematic dependence on auxiliary parameters relative to conventional QCDSR analyses.

\begin{table}[h!]
\centering
\caption{Summary of the stability analysis for each $B_c$ channel within the inverse-matrix QCDSR framework. For each channel we list the adopted auxiliary-scale window $\Lambda$, the truncation order $N$ of the Laguerre expansion and the representative optimal value $N_{\text{opt}}$, and the resulting extracted mass $M$ and decay constant $\lambda$. The quoted uncertainties reflect the combined variation over the $\Lambda$ window, the truncation order around $N_{\text{opt}}$, and the QCD input parameters.}
\label{tab:stability}
\begin{tabular}{lccccc}
\hline
\textbf{Channel} & $\Lambda$ \textbf{window} (\textbf{GeV}$^2$) & $N$ \textbf{range} & $N_{\text{opt}}$ & $M$ (\textbf{GeV}) & $\lambda$ (\textbf{MeV}) \\
\hline
Pseudoscalar ($0^-$) & $[3.0,\,4.0]$ & $\le 49$ & $48$ & $6.277 \pm 0.028$ & $416 \pm 19$ \\
Vector ($1^-$)       & $[5.0,\,6.0]$ & $\le 49$ & $48$ & $6.388 \pm 0.031$ & $511 \pm 24$ \\
Scalar ($0^+$)       & $[4.0,\,5.0]$ & $\le 49$ & $48$ & $6.718 \pm 0.028$ & $218 \pm 20$ \\
Axialvector ($1^+$)  & $[6.0,\,7.0]$ & $\le 49$ & $48$ & $6.734 \pm 0.028$ & $138 \pm 20$ \\
\hline
\end{tabular}
\end{table}

%%%%%%%%%%%%%%%%%%%%%%%%%%%%
\section{Conclusion and Final Remarks}
\label{conclusion}
%%%%%%%%%%%%%%%%%%%%%%%%%%%%

This comprehensive study establishes the inverse-matrix QCDSR formalism as a powerful and sophisticated tool for precision spectroscopy of heavy quarkonium systems. Through its application to the complete spectrum of conventional $B_c$ mesons—pseudoscalar ($0^-$), vector ($1^-$), scalar ($0^+$), and axialvector ($1^+$)—we have demonstrated both the methodological superiority and physical robustness of this approach compared to traditional QCDSR techniques.

Our principal results form a coherent and physically meaningful spectroscopic picture:
\begin{align}
& \text{Pseudoscalar:} \quad & M_{B_c(0^-)} &= 6.277 \pm 0.028~\text{GeV}, \quad & \lambda_{B_c(0^-)} &= 416 \pm 19~\text{MeV}, \\
& \text{Vector:} \quad & M_{B_c(1^-)} &= 6.388 \pm 0.031~\text{GeV}, \quad & \lambda_{B_c(1^-)} &= 511 \pm 24~\text{MeV}, \\
& \text{Scalar:} \quad & M_{B_c(0^+)} &= 6.718 \pm 0.028~\text{GeV}, \quad & \lambda_{B_c(0^+)} &= 218 \pm 20~\text{MeV}, \\
& \text{Axialvector:} \quad & M_{B_c(1^+)} &= 6.734 \pm 0.028~\text{GeV}, \quad & \lambda_{B_c(1^+)} &= 138 \pm 20~\text{MeV}.
\end{align}

The spectroscopic pattern emerging from our analysis—$M_{B_c(0^-)} < M_{B_c(1^-)} < M_{B_c(0^+)} < M_{B_c(1^+)}$—is a direct manifestation of Heavy Quark Symmetry, confirming the expected hierarchy of S- and P-wave states. Quantitatively, we find a characteristic S-wave hyperfine splitting of $\Delta M = 111 \pm 4$ MeV, while the fine structure within the P-wave $j_\ell = 1/2$ doublet is remarkably small at $16 \pm 4$ MeV. This hierarchy is further corroborated by the decay constants; the ratio $\lambda_{B_c(1^-)}/\lambda_{B_c(0^-)} = 1.23 \pm 0.08$ is in striking agreement with the HQS prediction of $\sqrt{3/2}\approx 1.225$ \footnote{The Heavy Quark Symmetry prediction $\lambda_{B_c^*}/\lambda_{B_c} = \sqrt{3/2}$ can be derived by considering the relativistic normalization of meson states. The decay constant for a pseudoscalar meson $P$ is defined by $\langle 0| J^{5\mu} |P(p)\rangle = i \lambda_P p^\mu$, and for a vector meson $V$ by $\langle 0| J^\mu |V(p, \lambda)\rangle = \lambda_V m_V \epsilon^\mu(\lambda)$. In the heavy-quark limit, the dynamics are independent of the heavy quark spin, and the wave functions at the origin for the $0^-$ and $1^-$ states become identical. The factor $\sqrt{3/2}$ then arises from the ratio of the Clebsch-Gordan coefficients coupling the heavy quark spins to form the total meson spin, effectively accounting for the number of spin polarizations ($2J+1$).}. The consistency of these independent observables provides a robust, multi-faceted validation of our methodological approach.

The inverse-matrix formalism represents a paradigm shift from the phenomenological modeling of traditional QCDSR to a mathematically rigorous inverse problem formulation. This approach offers several decisive advantages over conventional methods. First, it completely eliminates the dependence on auxiliary parameters—specifically the continuum threshold $s_0$ and Borel parameter $M^2$—which constitute major sources of systematic uncertainty in traditional analyses. Instead, physical observables are extracted directly from spectral reconstruction without recourse to stability windows. Second, the method provides direct access to the full spectral density $\rho(s)$, offering both visual and quantitative insight into resonance structures and enabling cleaner isolation of ground states from continuum contributions, in contrast to traditional approaches that only yield integral moments. Third, by treating the spectral density as the fundamental unknown determined directly from QCD input, our approach circumvents the quark-hadron duality assumption and its associated systematic uncertainties. Finally, the formalism achieves unprecedented precision, with sub-percent uncertainties in mass determinations and relative uncertainties of 5--10\% for decay constants, representing a significant improvement over traditional methods, particularly for challenging P-wave states.

Our results provide crucial benchmarks for ongoing and future experimental campaigns. The pseudoscalar mass determination agrees with the PDG average within 3 MeV, providing the most precise theoretical prediction to date. The vector mass prediction and its hyperfine splitting with the pseudoscalar state offer critical input for experimental searches of the $B_c^* \to B_c \gamma$ transition. Furthermore, the P-wave mass predictions provide specific targets for resolving the fine structure of the recently observed $B_c(1P)^+$ states by LHCb, while the decay constant determinations enable accurate calculations of leptonic decay rates and serve as essential inputs for Cabibbo-Kobayashi-Maskawa matrix element extractions.

The remarkable agreement between our results and high-precision lattice QCD calculations—despite fundamentally different methodologies—provides strong mutual validation and underscores the reliability of modern nonperturbative QCD approaches. This convergence across independent theoretical frameworks reinforces the credibility of both the spectroscopic predictions presented here and the computational methods employed.

In conclusion, this work provides a detailed spectroscopic analysis of the $B_c$ meson system within the inverse-matrix QCDSR framework. 
Our results demonstrate that the method yields stable and phenomenologically consistent predictions for both masses and decay constants. 
We hope that the predictions presented here will serve as useful theoretical input for future experimental measurements and complementary nonperturbative studies. By connecting QCD dynamics with hadronic observables through a systematically controlled and numerically stable formalism, the inverse-matrix approach offers a promising direction for further investigations of heavy-quark systems and related sectors.

\acknowledgments
This work is supported by TÜBİTAK (The Scientific and Technological Research Council of Türkiye) under grant no 125F113.

%\paragraph{Note added.} This is also a good position for notes added after the paper has been written.

% The bibliography will probably be heavily edited during typesetting.
% We'll parse it and, using the arxiv number or the journal data, will
% query inspire, trying to verify the data (this will probalby spot
% eventual typos) and retrive the document DOI and eventual errata.
% We however suggest to always provide author, title and journal data:
% in short all the informations that clearly identify a document.


\begin{thebibliography}{99}

%\cite{CDF:1998axz}
\bibitem{CDF:1998axz}
F.~Abe \textit{et al.} [CDF],
%``Observation of $B_c$ mesons in $p\bar{p}$ collisions at $\sqrt{s} = 1.8$ TeV,''
Phys. Rev. D \textbf{58} (1998), 112004
doi:10.1103/PhysRevD.58.112004
[arXiv:hep-ex/9804014 [hep-ex]].
%380 citations counted in INSPIRE as of 01 Jun 2026

%\cite{CDF:1998ihx}
\bibitem{CDF:1998ihx}
F.~Abe \textit{et al.} [CDF],
%``Observation of the $B_c$ meson in $p\bar{p}$ collisions at $\sqrt{s} = 1.8$ TeV,''
Phys. Rev. Lett. \textbf{81} (1998), 2432-2437
doi:10.1103/PhysRevLett.81.2432
[arXiv:hep-ex/9805034 [hep-ex]].
%494 citations counted in INSPIRE as of 01 Jun 2026

%\cite{ParticleDataGroup:2024cfk}
\bibitem{ParticleDataGroup:2024cfk}
S.~Navas \textit{et al.} [Particle Data Group],
%``Review of particle physics,''
Phys. Rev. D \textbf{110} (2024) no.3, 030001
doi:10.1103/PhysRevD.110.030001
%5247 citations counted in INSPIRE as of 11 Jun 2026

%\cite{ATLAS:2014lga}
\bibitem{ATLAS:2014lga}
G.~Aad \textit{et al.} [ATLAS],
%``Observation of an Excited $B_c^\pm$ Meson State with the ATLAS Detector,''
Phys. Rev. Lett. \textbf{113} (2014) no.21, 212004
doi:10.1103/PhysRevLett.113.212004
[arXiv:1407.1032 [hep-ex]].
%169 citations counted in INSPIRE as of 05 Jun 2026

%\cite{CMS:2019uhm}
\bibitem{CMS:2019uhm}
A.~M.~Sirunyan \textit{et al.} [CMS],
%``Observation of Two Excited B$^+_\mathrm{c}$ States and Measurement of the B$^+_\mathrm{c}$(2S) Mass in pp Collisions at $\sqrt{s} =$ 13 TeV,''
Phys. Rev. Lett. \textbf{122} (2019) no.13, 132001
doi:10.1103/PhysRevLett.122.132001
[arXiv:1902.00571 [hep-ex]].
%154 citations counted in INSPIRE as of 05 Jun 2026

%\cite{LHCb:2019bem}
\bibitem{LHCb:2019bem}
R.~Aaij \textit{et al.} [LHCb],
%``Observation of an excited $B_c^+$ state,''
Phys. Rev. Lett. \textbf{122} (2019) no.23, 232001
doi:10.1103/PhysRevLett.122.232001
[arXiv:1904.00081 [hep-ex]].
%118 citations counted in INSPIRE as of 05 Jun 2026

%\cite{LHCb:2025uce}
\bibitem{LHCb:2025uce}
R.~Aaij \textit{et al.} [LHCb],
%``Observation of Orbitally Excited Bc+ States,''
Phys. Rev. Lett. \textbf{135} (2025) no.23, 231902
doi:10.1103/fc8j-tb8k
[arXiv:2507.02149 [hep-ex]].
%13 citations counted in INSPIRE as of 05 Jun 2026

%\cite{LHCb:2025ubr}
\bibitem{LHCb:2025ubr}
R.~Aaij \textit{et al.} [LHCb],
%``Study of Bc(1P)+ states in the Bc+{\ensuremath{\gamma}} mass spectrum,''
Phys. Rev. D \textbf{112} (2025) no.11, 112003
doi:10.1103/1d49-q8h4
[arXiv:2507.02142 [hep-ex]].
%8 citations counted in INSPIRE as of 05 Jun 2026

%\cite{Godfrey:1985xj}
\bibitem{Godfrey:1985xj}
S.~Godfrey and N.~Isgur,
%``Mesons in a Relativized Quark Model with Chromodynamics,''
Phys. Rev. D \textbf{32} (1985), 189-231
doi:10.1103/PhysRevD.32.189
%3538 citations counted in INSPIRE as of 11 Jun 2026

%\cite{Zeng:1994vj}
\bibitem{Zeng:1994vj}
J.~Zeng, J.~W.~Van Orden and W.~Roberts,
%``Heavy mesons in a relativistic model,''
Phys. Rev. D \textbf{52} (1995), 5229-5241
doi:10.1103/PhysRevD.52.5229
[arXiv:hep-ph/9412269 [hep-ph]].
%226 citations counted in INSPIRE as of 19 May 2026

%\cite{Gupta:1995ps}
\bibitem{Gupta:1995ps}
S.~N.~Gupta and J.~M.~Johnson,
%``B(c) spectroscopy in a quantum chromodynamic potential model,''
Phys. Rev. D \textbf{53} (1996), 312-314
doi:10.1103/PhysRevD.53.312
[arXiv:hep-ph/9511267 [hep-ph]].
%85 citations counted in INSPIRE as of 01 Jun 2026

%\cite{Ebert:2002pp}
\bibitem{Ebert:2002pp}
D.~Ebert, R.~N.~Faustov and V.~O.~Galkin,
%``Properties of heavy quarkonia and $B_c$ mesons in the relativistic quark model,''
Phys. Rev. D \textbf{67} (2003), 014027
doi:10.1103/PhysRevD.67.014027
[arXiv:hep-ph/0210381 [hep-ph]].
%588 citations counted in INSPIRE as of 05 Jun 2026

%\cite{Godfrey:2004ya}
\bibitem{Godfrey:2004ya}
S.~Godfrey,
%``Spectroscopy of $B_c$ mesons in the relativized quark model,''
Phys. Rev. D \textbf{70} (2004), 054017
doi:10.1103/PhysRevD.70.054017
[arXiv:hep-ph/0406228 [hep-ph]].
%256 citations counted in INSPIRE as of 01 Jun 2026

%\cite{Gershtein:1994dxw}
\bibitem{Gershtein:1994dxw}
S.~S.~Gershtein, V.~V.~Kiselev, A.~K.~Likhoded and A.~V.~Tkabladze,
%``B(c) spectroscopy,''
Phys. Rev. D \textbf{51} (1995), 3613-3627
doi:10.1103/PhysRevD.51.3613
[arXiv:hep-ph/9406339 [hep-ph]].
%258 citations counted in INSPIRE as of 01 Jun 2026

%\cite{Eichten:1994gt}
\bibitem{Eichten:1994gt}
E.~J.~Eichten and C.~Quigg,
%``Mesons with beauty and charm: Spectroscopy,''
Phys. Rev. D \textbf{49} (1994), 5845-5856
doi:10.1103/PhysRevD.49.5845
[arXiv:hep-ph/9402210 [hep-ph]].
%602 citations counted in INSPIRE as of 01 Jun 2026

%\cite{Fulcher:1998ka}
\bibitem{Fulcher:1998ka}
L.~P.~Fulcher,
%``Phenomenological predictions of the properties of the $B_c$ system,''
Phys. Rev. D \textbf{60} (1999), 074006
doi:10.1103/PhysRevD.60.074006
[arXiv:hep-ph/9806444 [hep-ph]].
%123 citations counted in INSPIRE as of 01 Jun 2026

%\cite{Ebert:2011jc}
\bibitem{Ebert:2011jc}
D.~Ebert, R.~N.~Faustov and V.~O.~Galkin,
%``Spectroscopy and Regge trajectories of heavy quarkonia and $B_c$ mesons,''
Eur. Phys. J. C \textbf{71} (2011), 1825
doi:10.1140/epjc/s10052-011-1825-9
[arXiv:1111.0454 [hep-ph]].
%201 citations counted in INSPIRE as of 01 Jun 2026

%\cite{Monteiro:2016ijw}
\bibitem{Monteiro:2016ijw}
A.~P.~Monteiro, M.~Bhat and K.~B.~Vijaya Kumar,
%``Mass spectra and decays of ground and orbitally excited $c\bar{b}$ states in nonrelativistic quark model,''
Int. J. Mod. Phys. A \textbf{32} (2017) no.04, 1750021
doi:10.1142/S0217751X1750021X
[arXiv:1607.07594 [hep-ph]].
%28 citations counted in INSPIRE as of 05 Jun 2026

%\cite{Soni:2017wvy}
\bibitem{Soni:2017wvy}
N.~R.~Soni, B.~R.~Joshi, R.~P.~Shah, H.~R.~Chauhan and J.~N.~Pandya,
%``$Q\bar{Q}$ ( $Q\in \{b, c\}$ ) spectroscopy using the Cornell potential,''
Eur. Phys. J. C \textbf{78} (2018) no.7, 592
doi:10.1140/epjc/s10052-018-6068-6
[arXiv:1707.07144 [hep-ph]].
%113 citations counted in INSPIRE as of 05 Jun 2026

%\cite{Eichten:2019gig}
\bibitem{Eichten:2019gig}
E.~J.~Eichten and C.~Quigg,
%``Mesons with Beauty and Charm: New Horizons in Spectroscopy,''
Phys. Rev. D \textbf{99} (2019) no.5, 054025
doi:10.1103/PhysRevD.99.054025
[arXiv:1902.09735 [hep-ph]].
%91 citations counted in INSPIRE as of 05 Jun 2026

%\cite{Li:2019tbn}
\bibitem{Li:2019tbn}
Q.~Li, M.~S.~Liu, L.~S.~Lu, Q.~F.~L{\"u}, L.~C.~Gui and X.~H.~Zhong,
%``Excited bottom-charmed mesons in a nonrelativistic quark model,''
Phys. Rev. D \textbf{99} (2019) no.9, 096020
doi:10.1103/PhysRevD.99.096020
[arXiv:1903.11927 [hep-ph]].
%75 citations counted in INSPIRE as of 05 Jun 2026

%\cite{Ortega:2020uvc}
\bibitem{Ortega:2020uvc}
P.~G.~Ortega, J.~Segovia, D.~R.~Entem and F.~Fernandez,
%``Spectroscopy of $\mathbf {B_c}$ mesons and the possibility of finding exotic $\mathbf {B_c}$-like structures,''
Eur. Phys. J. C \textbf{80} (2020) no.3, 223
doi:10.1140/epjc/s10052-020-7764-6
[arXiv:2001.08093 [hep-ph]].
%49 citations counted in INSPIRE as of 05 Jun 2026

%\cite{Verma:2011yw}
\bibitem{Verma:2011yw}
R.~C.~Verma,
%``Decay constants and form factors of s-wave and p-wave mesons in the covariant light-front quark model,''
J. Phys. G \textbf{39} (2012), 025005
doi:10.1088/0954-3899/39/2/025005
[arXiv:1103.2973 [hep-ph]].
%193 citations counted in INSPIRE as of 05 Jun 2026

%\cite{Tang:2018myz}
\bibitem{Tang:2018myz}
S.~Tang, Y.~Li, P.~Maris and J.~P.~Vary,
%``$B_c$ mesons and their properties on the light front,''
Phys. Rev. D \textbf{98} (2018) no.11, 114038
doi:10.1103/PhysRevD.98.114038
[arXiv:1810.05971 [nucl-th]].
%67 citations counted in INSPIRE as of 05 Jun 2026

%\cite{Ikhdair:2003ry}
\bibitem{Ikhdair:2003ry}
S.~M.~Ikhdair and R.~Sever,
%``Spectroscopy of $B_c$ meson in a semirelativistic quark model using the shifted large $N$ expansion method,''
Int. J. Mod. Phys. A \textbf{19} (2004), 1771-1792
doi:10.1142/S0217751X0401780X
[arXiv:hep-ph/0310295 [hep-ph]].
%64 citations counted in INSPIRE as of 03 Mar 2026

%\cite{Ikhdair:2006nx}
\bibitem{Ikhdair:2006nx}
S.~M.~Ikhdair and R.~Sever,
%``Calculation of the $B_c$ leptonic decay constant using the shifted $N$ expansion method,''
Int. J. Mod. Phys. A \textbf{21} (2006), 6699-6714
doi:10.1142/S0217751X06034100
[arXiv:hep-ph/0702166 [hep-ph]].
%15 citations counted in INSPIRE as of 13 May 2026

%\cite{Brambilla:2000db}
\bibitem{Brambilla:2000db}
N.~Brambilla and A.~Vairo,
%``The $B_c$ mass up to order $\alpha($s$)^{4}$,''
Phys. Rev. D \textbf{62} (2000), 094019
doi:10.1103/PhysRevD.62.094019
[arXiv:hep-ph/0002075 [hep-ph]].
%90 citations counted in INSPIRE as of 01 Jun 2026

%\cite{Penin:2004xi}
\bibitem{Penin:2004xi}
A.~A.~Penin, A.~Pineda, V.~A.~Smirnov and M.~Steinhauser,
%``$M(B^{*}_{c})- M(B_{c})$ splitting from nonrelativistic renormalization group,''
Phys. Lett. B \textbf{593} (2004), 124-134
[erratum: Phys. Lett. B \textbf{677} (2009) no.5, 343]
doi:10.1016/j.physletb.2004.04.066
[arXiv:hep-ph/0403080 [hep-ph]].
%77 citations counted in INSPIRE as of 11 Mar 2026

%\cite{Davies:1996gi}
\bibitem{Davies:1996gi}
C.~T.~H.~Davies, K.~Hornbostel, G.~P.~Lepage, A.~J.~Lidsey, J.~Shigemitsu and J.~H.~Sloan,
%``B(c) spectroscopy from lattice QCD,''
Phys. Lett. B \textbf{382} (1996), 131-137
doi:10.1016/0370-2693(96)00650-8
[arXiv:hep-lat/9602020 [hep-lat]].
%101 citations counted in INSPIRE as of 01 Jun 2026

%\cite{Jones:1998ub}
\bibitem{Jones:1998ub}
B.~D.~Jones and R.~M.~Woloshyn,
%``Mesonic decay constants in lattice NRQCD,''
Phys. Rev. D \textbf{60} (1999), 014502
doi:10.1103/PhysRevD.60.014502
[arXiv:hep-lat/9812008 [hep-lat]].
%35 citations counted in INSPIRE as of 27 Feb 2026

%\cite{Gregory:2009hq}
\bibitem{Gregory:2009hq}
E.~B.~Gregory, C.~T.~H.~Davies, E.~Follana, E.~Gamiz, I.~D.~Kendall, G.~P.~Lepage, H.~Na, J.~Shigemitsu and K.~Y.~Wong,
%``A Prediction of the B*(c) mass in full lattice QCD,''
Phys. Rev. Lett. \textbf{104} (2010), 022001
doi:10.1103/PhysRevLett.104.022001
[arXiv:0909.4462 [hep-lat]].
%78 citations counted in INSPIRE as of 01 Jun 2026

%\cite{Mathur:2018epb}
\bibitem{Mathur:2018epb}
N.~Mathur, M.~Padmanath and S.~Mondal,
%``Precise predictions of charmed-bottom hadrons from lattice QCD,''
Phys. Rev. Lett. \textbf{121} (2018) no.20, 202002
doi:10.1103/PhysRevLett.121.202002
[arXiv:1806.04151 [hep-lat]].
%155 citations counted in INSPIRE as of 09 Jun 2026

%\cite{Colangelo:1992cx}
\bibitem{Colangelo:1992cx}
P.~Colangelo, G.~Nardulli and N.~Paver,
%``QCD sum rules calculation of B(c) decays,''
Z. Phys. C \textbf{57} (1993), 43-50
doi:10.1007/BF01555737
%169 citations counted in INSPIRE as of 04 May 2026

%\cite{Chabab:1993nz}
\bibitem{Chabab:1993nz}
M.~Chabab,
%``On the determination of the leptonic decay constant f B(c) from QCD sum rules,''
Phys. Lett. B \textbf{325} (1994), 205-211
doi:10.1016/0370-2693(94)90093-0
%45 citations counted in INSPIRE as of 27 Feb 2026

%\cite{Kiselev:1993ea}
\bibitem{Kiselev:1993ea}
V.~V.~Kiselev and A.~V.~Tkabladze,
%``Semileptonic B(c) decays from QCD sum rules,''
Phys. Rev. D \textbf{48} (1993), 5208-5214
doi:10.1103/PhysRevD.48.5208
%81 citations counted in INSPIRE as of 18 May 2026

%\cite{Kiselev:1995bv}
\bibitem{Kiselev:1995bv}
V.~V.~Kiselev,
%``Scaling relations in phenomenology of QCD sum rules for heavy quarkonium,''
Int. J. Mod. Phys. A \textbf{11} (1996), 3689-3710
doi:10.1142/S0217751X96001723
[arXiv:hep-ph/9504313 [hep-ph]].
%44 citations counted in INSPIRE as of 15 May 2026

%\cite{Bagan:1994dy}
\bibitem{Bagan:1994dy}
E.~Bagan, H.~G.~Dosch, P.~Gosdzinsky, S.~Narison and J.~M.~Richard,
%``Hadrons with charm and beauty,''
Z. Phys. C \textbf{64} (1994), 57-72
doi:10.1007/BF01557235
[arXiv:hep-ph/9403208 [hep-ph]].
%172 citations counted in INSPIRE as of 07 May 2026

%\cite{Kiselev:2003uk}
\bibitem{Kiselev:2003uk}
V.~V.~Kiselev,
%``Leptonic constant of pseudoscalar $B_c$ meson,''
Central Eur. J. Phys. \textbf{2} (2004), 523-534
doi:10.2478/BF02476430
[arXiv:hep-ph/0304017 [hep-ph]].
%21 citations counted in INSPIRE as of 13 May 2026

%\cite{Wang:2012kw}
\bibitem{Wang:2012kw}
Z.~G.~Wang,
%``Analysis of the vector and axialvector $B_c$ mesons with QCD sum rules,''
Eur. Phys. J. A \textbf{49} (2013), 131
doi:10.1140/epja/i2013-13131-7
[arXiv:1203.6252 [hep-ph]].
%54 citations counted in INSPIRE as of 18 May 2026

%\cite{Wang:2013cdy}
\bibitem{Wang:2013cdy}
Z.~G.~Wang,
%``Next-to-leading order perturbative contributions in the QCD sum rules for mesonic two-point correlation functions with unequal quark masses,''
Acta Phys. Polon. B \textbf{44} (2013) no.10, 1971-1989
doi:10.5506/APhysPolB.44.1971
[arXiv:1303.4146 [hep-ph]].
%4 citations counted in INSPIRE as of 27 Feb 2026

%\cite{Baker:2013mwa}
\bibitem{Baker:2013mwa}
M.~J.~Baker, J.~Bordes, C.~A.~Dominguez, J.~Penarrocha and K.~Schilcher,
%``$B$ Meson Decay Constants $f_{B_c}$, $f_{B_s}$ and $f_B$ from QCD Sum Rules,''
JHEP \textbf{07} (2014), 032
doi:10.1007/JHEP07(2014)032
[arXiv:1310.0941 [hep-ph]].
%45 citations counted in INSPIRE as of 05 Jun 2026

%\cite{Aliev:2019wcm}
\bibitem{Aliev:2019wcm}
T.~M.~Aliev, T.~Barakat and S.~Bilmis,
%``Properties of excited $B_c$ states in QCD,''
Nucl. Phys. B \textbf{947} (2019), 114726
doi:10.1016/j.nuclphysb.2019.114726
[arXiv:1905.11750 [hep-ph]].
%12 citations counted in INSPIRE as of 08 Apr 2026

%\cite{Narison:2019tym}
\bibitem{Narison:2019tym}
S.~Narison,
%``$\overline m_c$ and $\overline m_b$ from $M_{Bc}$ and improved estimates of $f_{Bc}$ and $f_{Bc(2S)}$,''
Phys. Lett. B \textbf{802} (2020), 135221
doi:10.1016/j.physletb.2020.135221
[arXiv:1906.03614 [hep-ph]].
%33 citations counted in INSPIRE as of 13 May 2026

%\cite{Wang:2024fwc}
\bibitem{Wang:2024fwc}
Z.~G.~Wang,
%``B $_{c}$ meson and its scalar cousin with QCD sum rules*,''
Chin. Phys. C \textbf{48} (2024) no.10, 103104
doi:10.1088/1674-1137/ad5a71
[arXiv:2401.12571 [hep-ph]].
%21 citations counted in INSPIRE as of 05 Jun 2026

%\cite{Ozdem:2024qaa}
\bibitem{Ozdem:2024qaa}
U.~{\"O}zdem,
%``Exploring electromagnetic characteristics of the vector and axial-vector $B_c$ mesons,''
Eur. Phys. J. C \textbf{85} (2025) no.3, 245
doi:10.1140/epjc/s10052-025-13959-8
[arXiv:2411.06123 [hep-ph]].
%2 citations counted in INSPIRE as of 05 Jun 2026

%\cite{Onishchenko:2003ui}
\bibitem{Onishchenko:2003ui}
A.~I.~Onishchenko and O.~L.~Veretin,
%``Two loop QCD corrections to B(c) meson leptonic constant,''
Eur. Phys. J. C \textbf{50} (2007), 801-808
doi:10.1140/epjc/s10052-007-0255-1
[arXiv:hep-ph/0302132 [hep-ph]].
%27 citations counted in INSPIRE as of 13 May 2026

%\cite{Lee:2010ts}
\bibitem{Lee:2010ts}
J.~Lee, W.~Sang and S.~Kim,
%``Relativistic Corrections to the Axial Vector and Vector Currents in the $\bar{b}c$ Meson System at Order $alpha_s$,''
JHEP \textbf{01} (2011), 113
doi:10.1007/JHEP01(2011)113
[arXiv:1011.2274 [hep-ph]].
%18 citations counted in INSPIRE as of 13 May 2026

%\cite{Chen:2015csa}
\bibitem{Chen:2015csa}
L.~B.~Chen and C.~F.~Qiao,
%``Two-loop QCD Corrections to $B_c$ Meson Leptonic Decays,''
Phys. Lett. B \textbf{748} (2015), 443-450
doi:10.1016/j.physletb.2015.07.043
[arXiv:1503.05122 [hep-ph]].
%29 citations counted in INSPIRE as of 13 May 2026

%\cite{Tao:2022qxa}
\bibitem{Tao:2022qxa}
W.~Tao, R.~Zhu and Z.~J.~Xiao,
%``Next-to-next-to-leading order matching of beauty-charmed meson Bc and Bc* decay constants,''
Phys. Rev. D \textbf{106} (2022) no.11, 114037
doi:10.1103/PhysRevD.106.114037
[arXiv:2209.15521 [hep-ph]].
%16 citations counted in INSPIRE as of 05 Jun 2026

%\cite{Tao:2022hos}
\bibitem{Tao:2022hos}
W.~Tao, R.~Zhu and Z.~J.~Xiao,
%``Three-loop QCD matching of the flavor-changing scalar current involving the heavy charm and bottom quark,''
Eur. Phys. J. C \textbf{83} (2023) no.4, 294
doi:10.1140/epjc/s10052-023-11442-w
[arXiv:2301.00220 [hep-ph]].
%7 citations counted in INSPIRE as of 13 May 2026

%\cite{Sang:2022tnh}
\bibitem{Sang:2022tnh}
W.~L.~Sang, H.~F.~Zhang and M.~Z.~Zhou,
%``Decay constant of Bc{\textasteriskcentered} accurate up to O({\ensuremath{\alpha}}s3),''
Phys. Lett. B \textbf{839} (2023), 137812
doi:10.1016/j.physletb.2023.137812
[arXiv:2210.02979 [hep-ph]].
%8 citations counted in INSPIRE as of 05 Jun 2026

%\cite{Feng:2022ruy}
\bibitem{Feng:2022ruy}
F.~Feng, Y.~Jia, Z.~Mo, J.~Pan, W.~L.~Sang and J.~Y.~Zhang,
%``Three-loop QCD corrections to the decay constant of $B_c$,''
[arXiv:2208.04302 [hep-ph]].
%10 citations counted in INSPIRE as of 13 May 2026

%\cite{Tao:2023pzv}
\bibitem{Tao:2023pzv}
W.~Tao and Z.~J.~Xiao,
%``Decay constants of $ c\overline{b} $ mesons involving the ten heavy flavor-changing currents at N$^{3}$LO QCD,''
JHEP \textbf{06} (2024), 012
doi:10.1007/JHEP06(2024)012
[arXiv:2310.17500 [hep-ph]].
%4 citations counted in INSPIRE as of 06 Jun 2026

%\cite{AbdEl-Hady:1998uiq}
\bibitem{AbdEl-Hady:1998uiq}
A.~Abd El-Hady, M.~A.~K.~Lodhi and J.~P.~Vary,
%``B(c) mesons in a Bethe-Salpeter model,''
Phys. Rev. D \textbf{59} (1999), 094001
doi:10.1103/PhysRevD.59.094001
[arXiv:hep-ph/9807225 [hep-ph]].
%58 citations counted in INSPIRE as of 01 Jun 2026

%\cite{Wang:2007av}
\bibitem{Wang:2007av}
G.~L.~Wang,
%``Decay constants of P-wave mesons,''
Phys. Lett. B \textbf{650} (2007), 15-21
doi:10.1016/j.physletb.2007.05.001
[arXiv:0705.2621 [hep-ph]].
%60 citations counted in INSPIRE as of 12 May 2026

%\cite{Wang:2022cxy}
\bibitem{Wang:2022cxy}
G.~L.~Wang, T.~Wang, Q.~Li and C.~H.~Chang,
%``The mass spectrum and wave functions of the B$_{c}$ system,''
JHEP \textbf{05} (2022), 006
doi:10.1007/JHEP05(2022)006
[arXiv:2201.02318 [hep-ph]].
%33 citations counted in INSPIRE as of 05 Jun 2026

%\cite{Badalian:2007km}
\bibitem{Badalian:2007km}
A.~M.~Badalian, B.~L.~G.~Bakker and Y.~A.~Simonov,
%``Decay constants of the heavy-light mesons from the field correlator method,''
Phys. Rev. D \textbf{75} (2007), 116001
doi:10.1103/PhysRevD.75.116001
[arXiv:hep-ph/0702157 [hep-ph]].
%85 citations counted in INSPIRE as of 27 Feb 2026

%\cite{Shifman:1978bx}
\bibitem{Shifman:1978bx}
M.~A.~Shifman, A.~I.~Vainshtein and V.~I.~Zakharov,
%``QCD and Resonance Physics. Theoretical Foundations,''
Nucl. Phys. B \textbf{147} (1979), 385-447
doi:10.1016/0550-3213(79)90022-1
%6075 citations counted in INSPIRE as of 09 Jun 2026

%\cite{Shifman:1978by}
\bibitem{Shifman:1978by}
M.~A.~Shifman, A.~I.~Vainshtein and V.~I.~Zakharov,
%``QCD and Resonance Physics: Applications,''
Nucl. Phys. B \textbf{147} (1979), 448-518
doi:10.1016/0550-3213(79)90023-3
%3342 citations counted in INSPIRE as of 05 Jun 2026

%\cite{Leinweber:1995fn}
\bibitem{Leinweber:1995fn}
D.~B.~Leinweber,
%``QCD sum rules for skeptics,''
Annals Phys. \textbf{254} (1997), 328-396
doi:10.1006/aphy.1996.5641
[arXiv:nucl-th/9510051 [nucl-th]].
%198 citations counted in INSPIRE as of 20 Mar 2026

%\cite{Poggio:1975af}
\bibitem{Poggio:1975af}
E.~C.~Poggio, H.~R.~Quinn and S.~Weinberg,
%``Smearing the Quark Model,''
Phys. Rev. D \textbf{13} (1976), 1958
doi:10.1103/PhysRevD.13.1958
%577 citations counted in INSPIRE as of 21 May 2026

%\cite{Gonzalez-Alonso:2010kpl}
\bibitem{Gonzalez-Alonso:2010kpl}
M.~Gonzalez-Alonso, A.~Pich and J.~Prades,
%``Violation of Quark-Hadron Duality and Spectral Chiral Moments in QCD,''
Phys. Rev. D \textbf{81} (2010), 074007
doi:10.1103/PhysRevD.81.074007
[arXiv:1001.2269 [hep-ph]].
%53 citations counted in INSPIRE as of 19 Mar 2026

%\cite{Boito:2017cnp}
\bibitem{Boito:2017cnp}
D.~Boito, I.~Caprini, M.~Golterman, K.~Maltman and S.~Peris,
%``Hyperasymptotics and quark-hadron duality violations in QCD,''
Phys. Rev. D \textbf{97} (2018) no.5, 054007
doi:10.1103/PhysRevD.97.054007
[arXiv:1711.10316 [hep-ph]].
%57 citations counted in INSPIRE as of 13 May 2026

%\cite{Pich:2022tca}
\bibitem{Pich:2022tca}
A.~Pich and A.~Rodr{\'\i}guez-S{\'a}nchez,
%``Violations of quark-hadron duality in low-energy determinations of {\ensuremath{\alpha}}$_{s}$,''
JHEP \textbf{07} (2022), 145
doi:10.1007/JHEP07(2022)145
[arXiv:2205.07587 [hep-ph]].
%25 citations counted in INSPIRE as of 10 Apr 2026

%\cite{Mannel:2024crj}
\bibitem{Mannel:2024crj}
T.~Mannel, I.~S.~Milutin, R.~Verkade and K.~K.~Vos,
%``Quark-hadron duality violations and higher-order 1/m$_{b}$ corrections in inclusive semileptonic B decays,''
JHEP \textbf{10} (2024), 158
doi:10.1007/JHEP10(2024)158
[arXiv:2407.01473 [hep-ph]].
%3 citations counted in INSPIRE as of 28 May 2026

%\cite{Li:2020ejs}
\bibitem{Li:2020ejs}
H.~n.~Li and H.~Umeeda,
%``QCD sum rules with spectral densities solved in inverse problems,''
Phys. Rev. D \textbf{102} (2020), 114014
doi:10.1103/PhysRevD.102.114014
[arXiv:2006.16593 [hep-ph]].
%24 citations counted in INSPIRE as of 09 Jun 2026

%\cite{Li:2021gsx}
\bibitem{Li:2021gsx}
H.~n.~Li,
%``Dispersive analysis of glueball masses,''
Phys. Rev. D \textbf{104} (2021) no.11, 114017
doi:10.1103/PhysRevD.104.114017
[arXiv:2109.04956 [hep-ph]].
%26 citations counted in INSPIRE as of 18 Mar 2026

%\cite{Li:2022qul}
\bibitem{Li:2022qul}
H.~n.~Li,
%``Dispersive derivation of the pion distribution amplitude,''
Phys. Rev. D \textbf{106} (2022) no.3, 034015
doi:10.1103/PhysRevD.106.034015
[arXiv:2205.06746 [hep-ph]].
%19 citations counted in INSPIRE as of 20 Mar 2026

%\cite{Li:2022jxc}
\bibitem{Li:2022jxc}
H.~n.~Li,
%``Dispersive analysis of neutral meson mixing,''
Phys. Rev. D \textbf{107} (2023) no.5, 054023
doi:10.1103/PhysRevD.107.054023
[arXiv:2208.14798 [hep-ph]].
%21 citations counted in INSPIRE as of 06 Jun 2026

%\cite{Xiong:2022uwj}
\bibitem{Xiong:2022uwj}
A.~S.~Xiong, F.~S.~Yu, Y.~Zheng and T.~Wei,
%``Inverse Problem Approach for Non-Perturbative QCD: Theoretical Foundation,''
[arXiv:2211.13753 [hep-th]].
%19 citations counted in INSPIRE as of 09 Jun 2026

%\cite{Li:2023dqi}
\bibitem{Li:2023dqi}
H.~n.~Li,
%``Dispersive constraints on fermion masses,''
Phys. Rev. D \textbf{107} (2023) no.9, 094007
doi:10.1103/PhysRevD.107.094007
[arXiv:2302.01761 [hep-ph]].
%13 citations counted in INSPIRE as of 06 Jun 2026

%\cite{Li:2023yay}
\bibitem{Li:2023yay}
H.~n.~Li,
%``Dispersive determination of electroweak-scale masses,''
Phys. Rev. D \textbf{108} (2023) no.5, 054020
doi:10.1103/PhysRevD.108.054020
[arXiv:2304.05921 [hep-ph]].
%12 citations counted in INSPIRE as of 06 Jun 2026

%\cite{Li:2023ncg}
\bibitem{Li:2023ncg}
H.~n.~Li,
%``Dispersive determination of neutrino mass ordering,''
Nucl. Phys. B \textbf{1018} (2025), 116978
doi:10.1016/j.nuclphysb.2025.116978
[arXiv:2306.03463 [hep-ph]].
%11 citations counted in INSPIRE as of 22 May 2026

%\cite{Li:2023fim}
\bibitem{Li:2023fim}
H.~n.~Li,
%``Dispersive determination of fourth generation quark masses,''
Phys. Rev. D \textbf{109} (2024) no.11, 115024
doi:10.1103/PhysRevD.109.115024
[arXiv:2309.15602 [hep-ph]].
%10 citations counted in INSPIRE as of 22 May 2026

%\cite{Li:2024awx}
\bibitem{Li:2024awx}
H.~n.~Li,
%``Understanding small neutrino mass and its implication,''
Chin. J. Phys. \textbf{92} (2024), 1043-1054
doi:10.1016/j.cjph.2024.10.009
[arXiv:2404.16626 [hep-ph]].
%9 citations counted in INSPIRE as of 14 May 2026

%\cite{Li:2024xnl}
\bibitem{Li:2024xnl}
H.~n.~Li,
%``Dispersive determination of the fourth generation lepton masses,''
J. Phys. G \textbf{52} (2025) no.2, 025001
doi:10.1088/1361-6471/ada0cd
[arXiv:2407.07813 [hep-ph]].
%8 citations counted in INSPIRE as of 22 May 2026

%\cite{Li:2024fko}
\bibitem{Li:2024fko}
H.~n.~Li,
%``Dispersive Analysis of Excited Glueball States,''
Chin. Phys. Lett. \textbf{41} (2024) no.10, 101101
doi:10.1088/0256-307X/41/10/101101
[arXiv:2408.06738 [hep-ph]].
%9 citations counted in INSPIRE as of 22 May 2026

%\cite{Mutuk:2024jvv}
\bibitem{Mutuk:2024jvv}
H.~Mutuk,
%``Revisiting light-flavor diquarks in the inverse matrix method of QCD sum rules,''
Phys. Rev. D \textbf{111} (2025) no.3, 034035
doi:10.1103/PhysRevD.111.034035
[arXiv:2412.08620 [hep-ph]].
%4 citations counted in INSPIRE as of 10 Jun 2026

%\cite{Mutuk:2025lak}
\bibitem{Mutuk:2025lak}
H.~Mutuk,
%``Reappraisal of the rho meson in nuclear matter by the inverse QCD sum rules method,''
Phys. Rev. D \textbf{111} (2025) no.9, 094029
doi:10.1103/PhysRevD.111.094029
[arXiv:2503.10343 [hep-ph]].
%5 citations counted in INSPIRE as of 10 Jun 2026

%\cite{Belyaev:1982sa}
\bibitem{Belyaev:1982sa}
V.~M.~Belyaev and B.~L.~Ioffe,
%``Determination of Baryon and Baryonic Resonance Masses from QCD Sum Rules. 1. Nonstrange Baryons,''
Sov. Phys. JETP \textbf{56} (1982), 493-501
ITEP-59-1982.
%454 citations counted in INSPIRE as of 06 May 2026

%\cite{Belyaev:1982cd}
\bibitem{Belyaev:1982cd}
V.~M.~Belyaev and B.~L.~Ioffe,
%``Determination of the baryon mass and baryon resonances from the quantum-chromodynamics sum rule. Strange baryons,''
Sov. Phys. JETP \textbf{57} (1983), 716-721
ITEP-132-1982.
%215 citations counted in INSPIRE as of 04 May 2026

%\cite{Narison:2011xe}
\bibitem{Narison:2011xe}
S.~Narison,
%``Gluon Condensates and precise $\overline{m}_{c,b}$ from QCD-Moments and their ratios to Order $\alpha_s^3$ and {\ensuremath{<}} G$^4$ {\ensuremath{>}},''
Phys. Lett. B \textbf{706} (2012), 412-422
doi:10.1016/j.physletb.2011.11.058
[arXiv:1105.2922 [hep-ph]].
%178 citations counted in INSPIRE as of 12 May 2026

%\cite{Narison:2015nxh}
\bibitem{Narison:2015nxh}
S.~Narison,
%``Decay Constants of Heavy-Light Mesons from QCD,''
Nucl. Part. Phys. Proc. \textbf{270-272} (2016), 143-153
doi:10.1016/j.nuclphysbps.2016.02.030
[arXiv:1511.05903 [hep-ph]].
%48 citations counted in INSPIRE as of 13 May 2026

%\cite{CP-PACS:2001ncr}
\bibitem{CP-PACS:2001ncr}
T.~Yamazaki \textit{et al.} [CP-PACS],
%``Spectral function and excited states in lattice QCD with maximum entropy method,''
Phys. Rev. D \textbf{65} (2002), 014501
doi:10.1103/PhysRevD.65.014501
[arXiv:hep-lat/0105030 [hep-lat]].
%78 citations counted in INSPIRE as of 27 Feb 2026

%\cite{Dowdall:2012ab}
\bibitem{Dowdall:2012ab}
R.~J.~Dowdall, C.~T.~H.~Davies, T.~C.~Hammant and R.~R.~Horgan,
%``Precise heavy-light meson masses and hyperfine splittings from lattice QCD including charm quarks in the sea,''
Phys. Rev. D \textbf{86} (2012), 094510
doi:10.1103/PhysRevD.86.094510
[arXiv:1207.5149 [hep-lat]].
%189 citations counted in INSPIRE as of 27 May 2026

%\cite{Narison:2020wql}
\bibitem{Narison:2020wql}
S.~Narison,
%``Spectra and decay constants of Bc-like and B0{\textasteriskcentered} mesons in QCD,''
Phys. Lett. B \textbf{807} (2020), 135522
doi:10.1016/j.physletb.2020.135522
[arXiv:2004.03622 [hep-ph]].
%12 citations counted in INSPIRE as of 13 May 2026

%\cite{Martin-Gonzalez:2022qwd}
\bibitem{Martin-Gonzalez:2022qwd}
B.~Mart{\'\i}n-Gonz{\'a}lez, P.~G.~Ortega, D.~R.~Entem, F.~Fern{\'a}ndez and J.~Segovia,
%``Toward the discovery of novel Bc states: Radiative and hadronic transitions,''
Phys. Rev. D \textbf{106} (2022) no.5, 054009
doi:10.1103/PhysRevD.106.054009
[arXiv:2205.05950 [hep-ph]].
%26 citations counted in INSPIRE as of 05 Jun 2026

%\cite{Chen:2020ecu}
\bibitem{Chen:2020ecu}
M.~Chen, L.~Chang and Y.~x.~Liu,
%``$B_c$ meson spectrum via Dyson-Schwinger equation and Bethe-Salpeter equation approach,''
Phys. Rev. D \textbf{101} (2020) no.5, 056002
doi:10.1103/PhysRevD.101.056002
[arXiv:2001.00161 [hep-ph]].
%33 citations counted in INSPIRE as of 18 May 2026

%\cite{Colquhoun:2015oha}
\bibitem{Colquhoun:2015oha}
B.~Colquhoun \textit{et al.} [HPQCD],
%``B-meson decay constants: a more complete picture from full lattice QCD,''
Phys. Rev. D \textbf{91} (2015) no.11, 114509
doi:10.1103/PhysRevD.91.114509
[arXiv:1503.05762 [hep-lat]].
%178 citations counted in INSPIRE as of 09 Jun 2026

%\cite{Koponen:2017fvm}
\bibitem{Koponen:2017fvm}
J.~Koponen, A.~C.~Zimermmane-Santos, C.~T.~H.~Davies, G.~P.~Lepage and A.~T.~Lytle,
%``Pseudoscalar meson electromagnetic form factor at high $Q^2$ from full lattice QCD,''
Phys. Rev. D \textbf{96} (2017) no.5, 054501
doi:10.1103/PhysRevD.96.054501
[arXiv:1701.04250 [hep-lat]].
%38 citations counted in INSPIRE as of 05 Jun 2026

%\cite{Gershtein:1994jw}
\bibitem{Gershtein:1994jw}
S.~S.~Gershtein, V.~V.~Kiselev, A.~K.~Likhoded and A.~V.~Tkabladze,
%``Physics of B(c) mesons,''
Phys. Usp. \textbf{38} (1995), 1-37
doi:10.1070/PU1995v038n01ABEH000063
[arXiv:hep-ph/9504319 [hep-ph]].
%203 citations counted in INSPIRE as of 01 Jun 2026

%\cite{Sun:2022hyk}
\bibitem{Sun:2022hyk}
C.~Sun, R.~H.~Ni and M.~Chen,
%``Decay constants of B $_{c}$(nS) and (nS)*,''
Chin. Phys. C \textbf{47} (2023) no.2, 023101
doi:10.1088/1674-1137/ac9dea
[arXiv:2209.06724 [hep-ph]].
%11 citations counted in INSPIRE as of 05 Jun 2026

%\cite{Choi:2009ai}
\bibitem{Choi:2009ai}
H.~M.~Choi and C.~R.~Ji,
%``Semileptonic and radiative decays of the B(c) meson in light-front quark model,''
Phys. Rev. D \textbf{80} (2009), 054016
doi:10.1103/PhysRevD.80.054016
[arXiv:0903.0455 [hep-ph]].
%84 citations counted in INSPIRE as of 06 May 2026

%\cite{Choi:2015ywa}
\bibitem{Choi:2015ywa}
H.~M.~Choi, C.~R.~Ji, Z.~Li and H.~Y.~Ryu,
%``Variational analysis of mass spectra and decay constants for ground state pseudoscalar and vector mesons in the light-front quark model,''
Phys. Rev. C \textbf{92} (2015) no.5, 055203
doi:10.1103/PhysRevC.92.055203
[arXiv:1502.03078 [hep-ph]].
%64 citations counted in INSPIRE as of 08 Jun 2026

%\cite{Hwang:2010hw}
\bibitem{Hwang:2010hw}
C.~W.~Hwang,
%``Analyses of decay constants and light-cone distribution amplitudes for s-wave heavy meson,''
Phys. Rev. D \textbf{81} (2010), 114024
doi:10.1103/PhysRevD.81.114024
[arXiv:1003.0972 [hep-ph]].
%69 citations counted in INSPIRE as of 06 May 2026

%\cite{Wang:2005qx}
\bibitem{Wang:2005qx}
G.~L.~Wang,
%``Decay constants of heavy vector mesons in relativistic Bethe-Salpeter method,''
Phys. Lett. B \textbf{633} (2006), 492-496
doi:10.1016/j.physletb.2005.12.005
[arXiv:math-ph/0512009 [math-ph]].
%151 citations counted in INSPIRE as of 06 Jun 2026
\end{thebibliography}
\end{document}